\documentclass[aps,prl,twocolumn,groupedaddress,showpacs,floatfix]{revtex4}
\usepackage{dcolumn}
\usepackage{amsmath}
\usepackage{amssymb}
\usepackage{graphicx}
\usepackage{bm}
\usepackage[T1]{fontenc}
\usepackage{color}
\usepackage{url}
\usepackage[bf]{subfigure}
\usepackage{rotating}

\usepackage{scalefnt}

\makeatletter
\newif\if@restonecol
\makeatother

\begin{document}

\title{{A process of rumor scotching on finite populations}}

\author{Guilherme Ferraz de Arruda}
\affiliation{Departamento de Matem\'{a}tica Aplicada e Estat\'{i}stica, Instituto de Ci\^{e}ncias Matem\'{a}ticas e de Computa\c{c}\~{a}o,
Universidade de S\~{a}o Paulo - Campus de S\~{a}o Carlos, Caixa Postal 668,13560-970 S\~{a}o Carlos, SP, Brazil.}

\author{Elcio Lebensztayn}
\affiliation{Instituto de Matem\'atica, Estat\'istica e Computa\c{c}\~ao Cient\'ifica, Universidade Estadual de Campinas -- UNICAMP, Rua S\'ergio Buarque de Holanda 651, CEP 13083-859, Campinas, SP, Brazil.}

\author{Francisco A. Rodrigues}
\affiliation{Departamento de Matem\'{a}tica Aplicada e Estat\'{i}stica, Instituto de Ci\^{e}ncias Matem\'{a}ticas e de Computa\c{c}\~{a}o,
Universidade de S\~{a}o Paulo - Campus de S\~{a}o Carlos, Caixa Postal 668,13560-970 S\~{a}o Carlos, SP, Brazil.}

\author{Pablo Mart\'{i}n Rodr\'{i}guez}
\email{pablor@icmc.usp.br}
\affiliation{Departamento de Matem\'{a}tica Aplicada e Estat\'{i}stica, Instituto de Ci\^{e}ncias Matem\'{a}ticas e de Computa\c{c}\~{a}o,
Universidade de S\~{a}o Paulo - Campus de S\~{a}o Carlos, Caixa Postal 668,13560-970 S\~{a}o Carlos, SP, Brazil.}

\begin{abstract}
Rumor spreading is a ubiquitous phenomenon in social and technological networks. Traditional models consider that the rumor is propagated by pairwise interactions between spreaders and ignorants. Spreaders can become stiflers only after contacting spreaders or stiflers. Here we propose a model that considers the traditional assumptions, but stiflers are active and try to scotch the rumor to the spreaders. An analytical treatment based on the theory of convergence of density dependent Markov chains is developed to analyze how the final proportion of ignorants behaves asymptotically in a finite homogeneously mixing population. We perform Monte Carlo simulations in random graphs and scale-free networks and verify that the results obtained for homogeneously mixing populations can be approximated for random graphs, but are not suitable for scale-free networks. Furthermore, regarding the process on a heterogeneous mixing population, we obtain a set of differential equations that describes the time evolution of the probability that an individual is in each state. Our model can be applied to study systems in which informed agents try to stop the rumor propagation. In addition, our results can be considered to develop optimal information dissemination strategies and approaches to control rumor propagation.
\end{abstract}

\maketitle

\section{Introduction}

Spreading phenomena is ubiquitous in nature and technology~\cite{Castellano09}. Diseases propagate from person to person, viruses contaminate computers worldwide and innovation spreads from place to place. In the last decades, the analysis of the phenomenon of information transmission from a mathematical and physical point of view has attracted the attention of many researchers~\cite{Keeling05,Hethcote00SR, Keeling08, Castellano09}. The expression ``information transmission'' is often used to refer to the spreading of news or rumors in a population or the diffusion of a virus through the Internet. These stochastic processes have similar properties and are often modeled by the same mathematical models~\cite{Keeling05,Hethcote00SR, Keeling08}. 

In this paper we propose and analyze a process of rumor scotching on finite populations. A removal mechanism different from the one considered in the usual models is considered here. i.e. we assume that stifler nodes can scotch the rumor propagation. Our model is inspired by the stochastic process discussed in~\cite{Bordenave/2008}. In such work, the author assumes that the propagation of a rumor starts from one individual, who after an exponential time learns that the rumor is false and then starts to scotch the propagation by the individuals previously informed. When the population is homogeneously mixed, Bordenave~\cite{Bordenave/2008} showed that the scaling limit of this process is the well-known birth-and-assassination process, introduced in the probabilistic literature by Aldous and Krebs~\cite{aldous/krebs/1990} as a variant of the branching process~\cite{athreya/ney/1972}. In order to introduce a more realistic model we consider two modifications. We suppose that each stifler tries to stop the rumor diffusion by all the spreaders that he/she meets along the way. It is assumed that the rumor starts with general initial conditions. An interacting particle system is considered to represent the spreading of the rumor by agents on a given graph. Then we assume that each agent may be in any of the three states belonging to the set $\{0,1,2\}$, where $0$ stands for ignorant, $1$ for spreader and $2$ for stifler. Finally, the model is formulated by considering that a spreader tells the rumor to any of its (nearest) ignorant neighbors at rate $\lambda$. A spreader becomes a stifler due to the action of its (nearest neighbor) stifler nodes at rate $\alpha$. 

Our model can be applied to describe the spreading of information through social networks. In this case, a person propagates a piece of information to another one and then becomes a stifler. After that, such person discovers that the piece of information is false and then tries to scotch the spreading. The same dynamics can model the spreading of data in a network. A computer can try to scotch the diffusion of a file after discovering that it contains a virus. This dynamics is related to the well-known Williams-Bjerknes (WB) tumor growth model~\cite{williams/bjerknes/1972}, which is studied on infinite regular graphs like hypercubic lattices and trees (see for instance~\cite{bramson1980,bramson1981,louidor}). The same model on complete graphs is studied by Kortchemski \cite{kortchemski/2014} in the context of a predator-prey SIR model. As a description of a rumor dynamic on finite graphs, including random graphs and scale-free networks, this model has not been addressed yet. In this way, here we apply the theory of convergence of density dependent Markov chains and use computational simulations to study rumor scotching on finite populations. The asymptotic behavior of the process in a homogeneously mixing population is analyzed. In addition, we simulate this model in complex networks in order to verify the cases in which the homogeneously mixing approximation is suitable. Furthermore, regarding the process on a heterogeneous mixing population, we obtain a set of differential equations that describes the time evolution of the probability that an individual is in each state. We show that there is a remarkable matching between these analytical results and those obtained from computer simulations. Our results can contribute to the analysis of optimal information dissemination strategies~\cite{Kandhway2014} as well as the statistical inference of rumor processes~\cite{leiva}.  

\section{Previous works on rumor spreading}

The most popular models to describe the spreading of news or rumors are based on the stochastic or deterministic version of the classical SIR (susceptible-infected-recovered), SIS (susceptible-infected-susceptible) and SI (susceptible-infected) epidemic models~\cite{Castellano09, PhysRevE.90.032812}. In these models, it is assumed that an infection (or information) spreads through a population subdivided into three classes (or compartments), i.e. susceptible, infective and removed individuals. In the case of rumor dynamics, these states are referred as ignorant, spreader and stifler, respectively.

The first stochastic rumor models are due to Daley and Kendall (DK)~\cite{daley/kendall/1964,daley/kendall/1965} and to Maki and Thompson (MT) \cite{maki/thompson/1973}. Both models were proposed to describe the diffusion of a rumor through a closed homogeneously mixing population of size $n$, i.e. a population described by a complete graph. Initially, it is assumed that there is one spreader and $n-1$ are in the ignorant state. The evolution of the DK rumor model can be described by using a continuous time Markov chain, denoting the number of nodes in the ignorant, spreader and stifler states at time $t$ by $X(t)$, $Y(t)$ and $Z(t)$, respectively. Thus, the stochastic process  $\{(X(t), Y(t))\}_{t \geq 0}$ is described by the Markov chain with transitions and corresponding rates given by
\begin{equation*}
{\allowdisplaybreaks
\begin{array}{cc}
\text{transition} \quad &\text{rate} \\[0.1cm]
(-1, 1) \quad &X Y, \\[0.1cm]
(0, -2) \quad &\displaystyle\binom{Y}{2}, \\[0.2cm]
(0, -1) \quad &Y (n  - X - Y).
\end{array}}%
\end{equation*}
This means that if the process is in state $(X,Y)$ at time $t$, then the probability that it will be in state $(X-1, Y+1)$ at time $t+h$ is $XY h+ o(h)$, where $o(h)$ is a function such that $\lim_{h \rightarrow 0}o(h)/h = 0$. In this model, it is assumed that individuals interact by pairwise contacts and the three possible transitions correspond to spreader-ignorant, spreader-spreader and spreader-stifler interactions. In the first transition, the spreader tells the rumor to an ignorant, who becomes a spreader. The two other transitions indicate the transformation of the spreader(s) into stifler(s) due to its contact with a subject who already knew the rumor. 

Maki and Thompson formulated a simplification of the DK model by considering that the rumor is propagated by directed contact between the spreaders and other individuals. In addition, when a spreader $i$ contacts another spreader $j$, only $i$ becomes a stifler. Thus, in this case, the continuous-time Markov chain to be considered is the stochastic process $\{(X(t), Y(t))\}_{t \geq 0}$ that evolves according to the following transitions and rates
\begin{equation*}
\begin{array}{cc}
\text{transition} \quad &\text{rate} \\[0.1cm]
(-1, 1) \quad &X Y, \\[0.1cm]
(0, -1) \quad &Y (n - X).
\end{array}
\end{equation*}
The first references about these models,~\cite{daley/kendall/1964,daley/kendall/1965,maki/thompson/1973}, are the most cited works about stochastic rumor processes in homogeneously mixing populations and have triggered  numerous significant research in this area. Basically, generalizations of these models can be obtained in two different ways. The first generalizations are related to the dynamic of the spreading process and the second ones to the structure of the population. In the former, there are many rigorous results involving the analysis of the remaining proportion of ignorant individuals when there are no more spreaders on the population~\cite{sudbury/1985,watson/1987}. Note that this is one way to measure the range of the rumor. After the first rigorous results, namely limit theorems for this fraction of ignorant individuals~\cite{sudbury/1985,watson/1987}, many authors introduced modifications in the dynamic of the basic models in order to make them more realistic. Recent papers have suggested generalizations that allow various contact interactions, the possibility of forgetting the rumor~\cite{kawachi/2008}, long-memory spreaders \cite{lebensztayn/machado/rodriguez/2011a}, or a new class of uninterested individuals \cite{lebensztayn/machado/rodriguez/2011b}. Related processes can be found for instance in~\cite{kurtz/lebensztayn/leichsenring/machado/2008,comets/delarue/schott/2013}. However, all these models maintain the assumption that the population is homogeneously mixing.

On the other hand, recent results have analyzed how the topology of the considered population affects the diffusion process. In this direction, Coletti et al.~\cite{coletti/schinazi/rodriguez/2012} studied a rumor process when the population is represented by the $d$-dimensional hypercubic lattice and Comets et al.~\cite{comets/gallesco/popov/vachkovskaia/2013} modeled the transmission of information of a message on the Erd\H{o}s-R\'enyi random graph. Related studies can be found in~\cite{berger/borgs/chayes/saberi/2005,durrett/jung/2007,junior/machado/zuluaga/2011,bertacchi/zucca/2013,gallo/garcia/junior/rodriguez/2014} and references therein. In the previous works, authors dealt with different probabilistic techniques to get the desired results. Such techniques allow extending our understanding of a rumor process in a more structured population, namely, represented by lattices and random graphs. Unfortunately, when one deals with the analysis of these dynamics in real-world networks, such as on-line social networks or the Internet~\cite{Costa011:AP}, whose topology is very heterogeneous, it is difficult to apply the same mathematical arguments and a different approach is required. In this direction, general rumor models are studied in~\cite{moreno/nekovee/pacheco/2004,isham/harden/nekovee/2010} where the population is represented by a random graph or a complex network and important results are obtained by means of approximations of the original process and computational simulations. 

\section{Homogeneously mixing populations} 
\label{sec:mf}

\begin{figure}[!t]
\begin{center}
\subfigure[][Variation of the parameters $\alpha$ and $\lambda$ for the fixed initial condition $x_0 = 0.98, y_0 = z_0 = 0.01$.]{\includegraphics[width=0.9\linewidth]{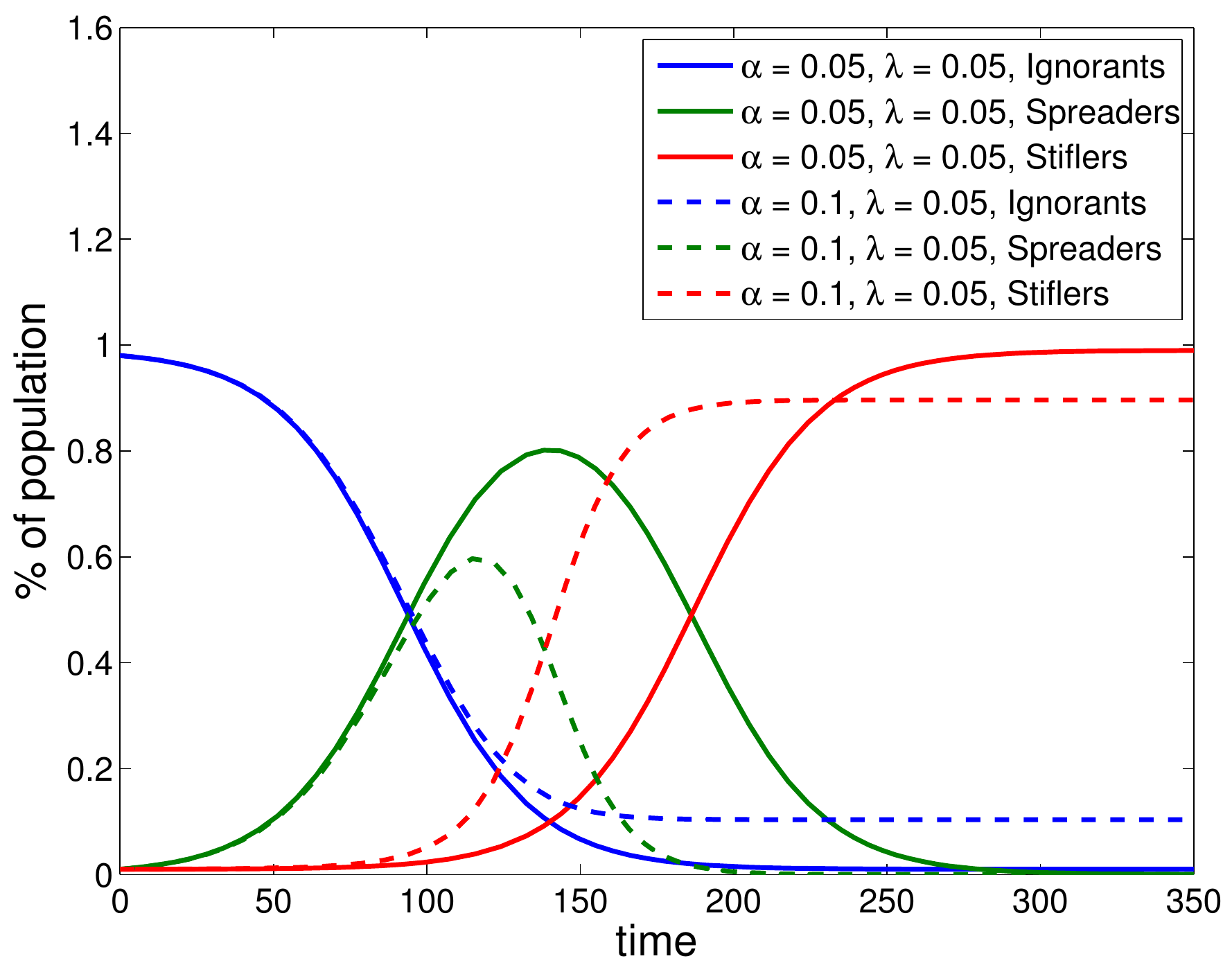}}
\qquad
\subfigure[][Variation of the initial condition for the fixed parameters $\alpha = 0.05, \lambda = 0.05$.]{\includegraphics[width=0.9\linewidth]{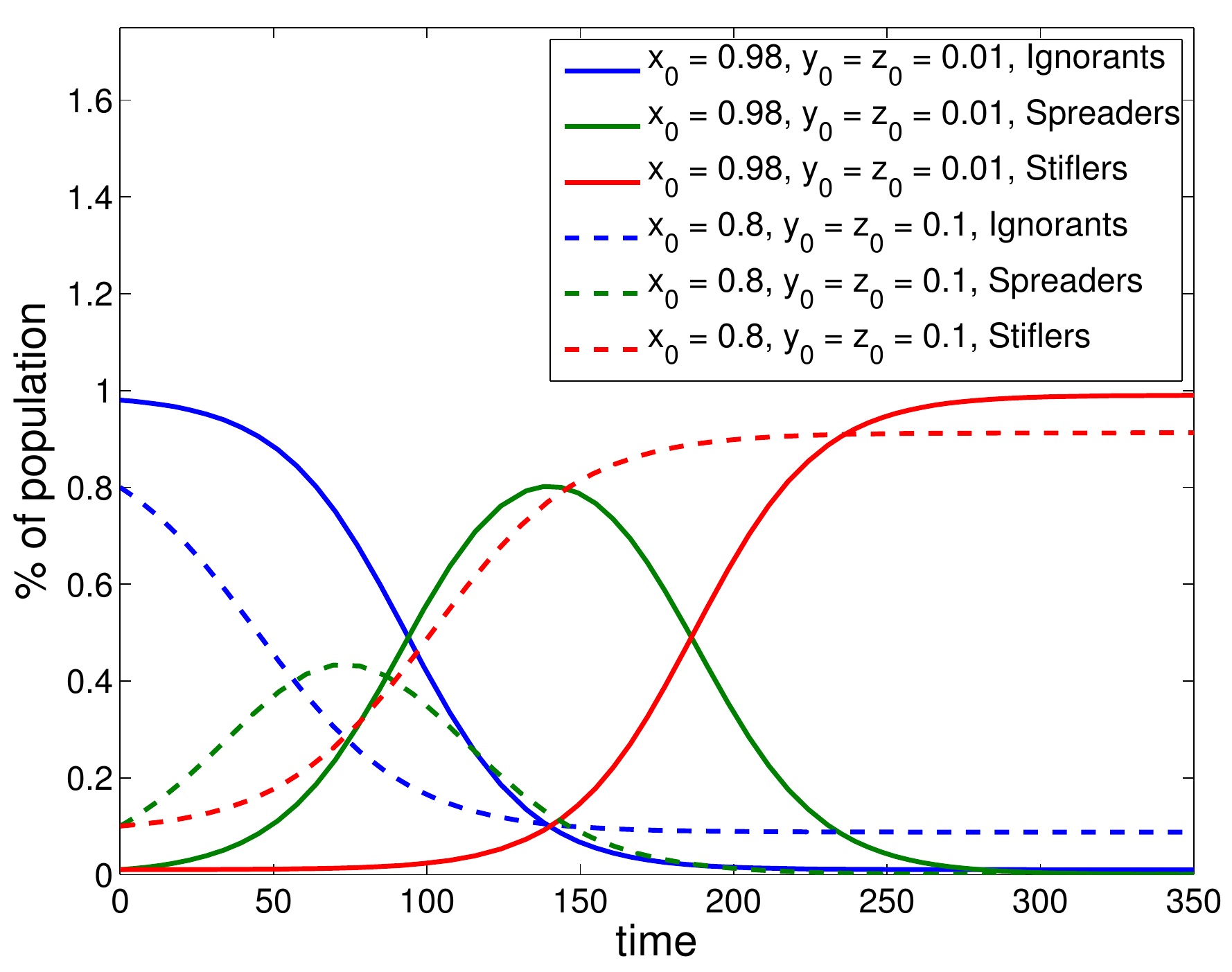}}
\end{center}
\caption{Time evolution of the rumor model (Eq.~\eqref{eq:edo})  according to the variation of (a) parameters $\alpha$ and $\lambda$,  or (b) initial condition.}
\label{fig:time}
\end{figure}

The model proposed here assumes that spreaders propagate the rumor to their direct neighbors, as in the original Maki-Thompson model~\cite{maki/thompson/1973}. However, differently from this model, stifler nodes try to scotch the rumor propagation. The corresponding dynamical process can be described by a set of differential equations given by
\begin{equation}
\begin{cases}
x^{\prime}(t) = - \lambda x(t) y(t), \\[0.1cm]
y^{\prime}(t) = \lambda x(t) y(t) - \alpha y(t) z(t), \\[0.1cm]
z^{\prime}(t) = \alpha y(t) z(t), \\[0.1cm]
x(0) = x_0, y(0) = y_0, z(0) = z_0,
\end{cases}
\label{eq:edo}
\end{equation}
where $x(t)$, $y(t)$ and $z(t)$ are the fractions of ignorant, spreader and stifler nodes at time $t$, respectively. We assume that a spreader tells the rumor to an ignorant at rate $\lambda$ and a spreader becomes a stifler at rate $\alpha$ due to the action of a stifler. The solutions rely on the initial conditions, since the stifler class is an absorbing state. Figure~\ref{fig:time} shows this dependency. In Figure~\ref{fig:time}(a), the initial conditions are fixed and two parameters $\alpha$ and $\lambda$ are evaluated, showing that an increase on the values of $\alpha$ reduces the maximum fraction of spreader nodes. In Figure~\ref{fig:time} (b), the rates are fixed and the initial conditions are varied, which shows that the time evolution of the system changes, evidencing the dependency on the initial conditions.

\begin{figure}[!t]
\begin{center}
\includegraphics[width=1\linewidth]{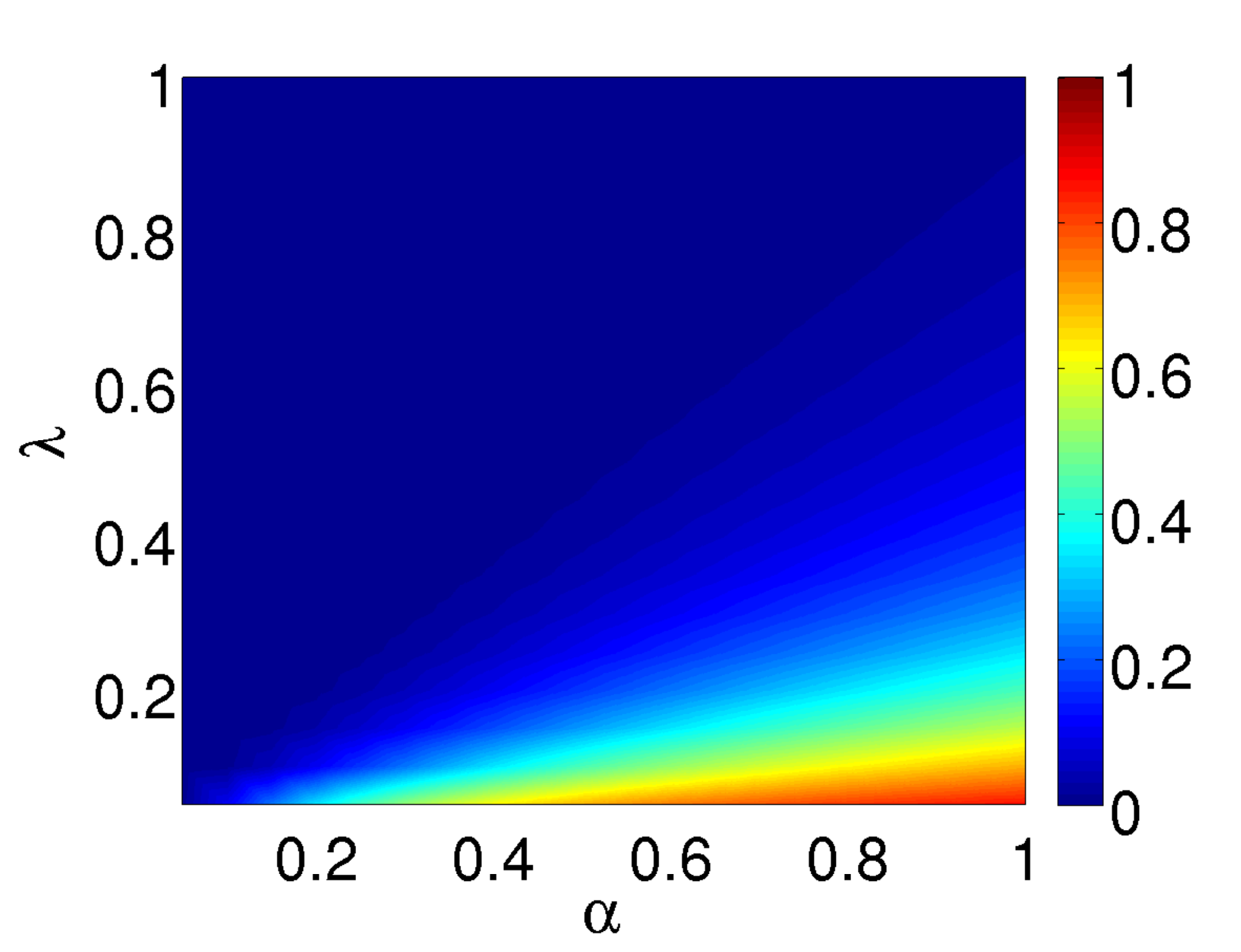}
\end{center}
\caption{Fraction of ignorant individuals for the theoretical model, obtained by the numerical evaluation of the system of Eqs.~\eqref{eq:edo} for $x_0 = 0.98$, $y_0 =
0.01$ and $z_0 = 0.01$.}
\label{fig:theoretical}
\end{figure}

The set of Eqs.~\eqref{eq:edo} describes the homogeneously mixing population assumption, in which every agent interacts with all the others with the same probability (mean-field approach). We solved this system numerically for every pair of parameters, $\lambda$ and $\alpha$, each one starting from 0.05 and incrementing them with steps of 0.05 until reaching the unity. Figure~\ref{fig:theoretical} (a) presents the results in terms of the fraction of ignorants at the end of the process. The higher the probability $\alpha$, the higher the fraction of the ignorants for low values of $\lambda$. On the other hand, the fraction of ignorants is lower when the parameter $\lambda$ is increased, even when $\alpha \approx 1$.

The homogeneously mixing population assumption (Eqs.~\eqref{eq:edo}) allows us to obtain some information about the remaining proportion of ignorants at the end of the process. However, this procedure refers to the limit of the process and it does not say us anything about the relation between such value and the size of the population. In order to study such relation we define a Markov chain to describe the process proposed. More specifically, we consider the theory of density dependent Markov chains, from which we can obtain not only information of the remaining proportion of ignorants, but also acquire a better understanding of the magnitude of the random fluctuations around this limiting value. This approach has already been used for rumor models, see for instance \cite{lebensztayn/machado/rodriguez/2011a,lebensztayn/machado/rodriguez/2011b}. 

Let us formalize the stochastic process of interest. Consider a population of fixed size $n$. As usual, we denote the number of nodes in the ignorant, spreaders and stiflers at time $t$ by $X^n(t)$, $Y^n(t)$ and $Z^n(t)$, respectively. We assume that $x_0^n$, $y_0^n$ and $z_0^n$ are the respective initial proportions of these individuals in the population and suppose that the following limits exist,
\begin{equation} \label{limites}
\begin{split}
x_0 &:= \lim_{n \to \infty} x^n_0 > 0;\\
y_0 &:= \lim_{n \to \infty} y^n_0 \quad \text{and} \\
z_0 &:= \lim_{n \to \infty} z^n_0 >0.
\end{split}
\end{equation}
Our rumor model is the continuous-time Markov chain $V^{(n)}(t)=\{(X^n(t), Y^n(t))\}_{t \geq 0}$ with transitions and rates are given by
\begin{equation*}
{\allowdisplaybreaks
\begin{array}{cc}
\text{transition} \quad &\text{rate} \\[0.1cm]
(-1, 1) \quad &\lambda X Y, \\[0.1cm]
(0, -1) \quad &\alpha Y (n - X - Y).
\end{array}}
\end{equation*}
This means that if the process is in state $(X,Y)$ at time $t$ then the probabilities that it will be in states $(X-1,Y+1)$ or $(X,Y-1)$ at time $t+h$ are, respectively, $\lambda\, X\,Y\,h + o(h)$ and $\alpha\, Y (n - X - Y)\,h + o(h)$. Note that while the first transition corresponds to an interaction between a spreader and an ignorant, the second one represents the interaction between a stifler and a spreader. When $n$ goes to infinity, the entire trajectories of this Markov chain have as a limit the set of differential equations in~\eqref{eq:edo}. 
In the rest of the paper, we denote the ratio $\alpha/\lambda$ by $\rho$. 

Thus defined, this model is an instance of the general stochastic rumor model proposed in \cite{lebensztayn/machado/rodriguez/2011b} with the choice of parameters given by $\delta = 1$, $\theta_1 = \theta_2 = 0$ and $\gamma = \rho$. However, following the notation used in~\cite{lebensztayn/machado/rodriguez/2011b}, 
$$
\theta = \theta_1 + \theta_2 - \gamma = - \rho < 0
$$
and the results obtained in that work cannot be applied directly. Nevertheless, the arguments presented here are quite similar. Let $ \tau^{(n)}= \inf \{t: Y^n(t) = 0 \}$ be the absorption time of the process. More specifically, $\tau^{(n)}$ is the first time at which the number of spreaders in the population vanishes. Our purpose is to study the behavior of the random variable $X^n (\tau^{(n)})/n$, for $n$ large enough, by stating a weak law of large numbers and a central limit theorem.

The main idea is to define, by means of a random time change, a new process $ \{ \tilde V^{(n)}(t) \}_{t \geq 0} $, with the same transitions as $ \{ V^{(n)}(t) \}_{t \geq 0} $, so that they terminate at the same point. The transformation is done in such a way that $ \{ \tilde V^{(n)}(t) \}_{t \geq 0} $ is a density dependent Markov chain for which we can apply well-known convergence results (see for instance \cite{Andersson/Britton/2000,Draief/Massouli/2010,Ethier2009}). 

\begin{figure*}[!t]
\begin{center}
\subfigure[][$\rho < \frac{x_0}{z_0}$ and $y_0 > 0$.]{\includegraphics[width=0.23\linewidth]{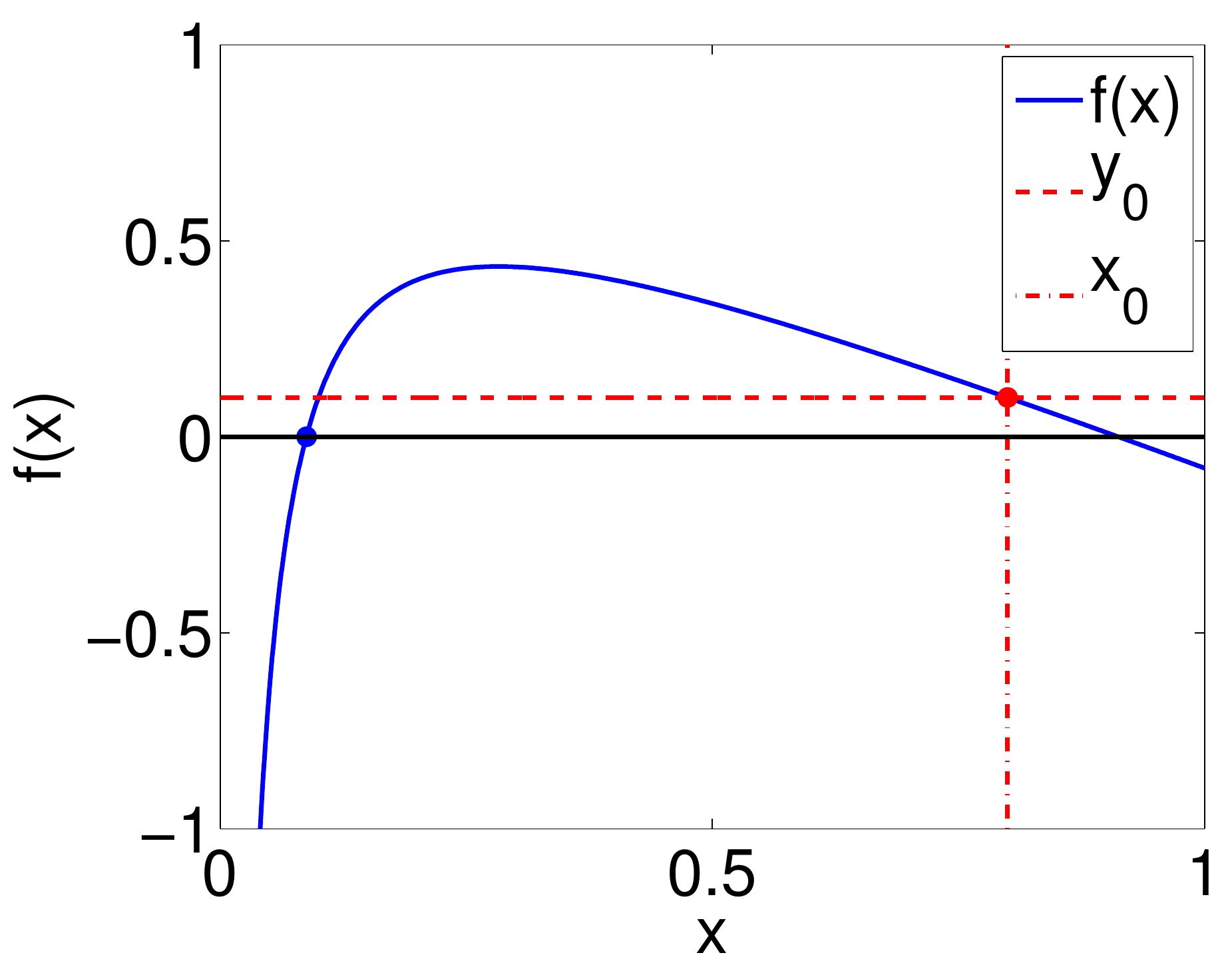}}
\subfigure[][$\rho > \frac{x_0}{z_0}$ and $y_0 > 0$.]{\includegraphics[width=0.23\linewidth]{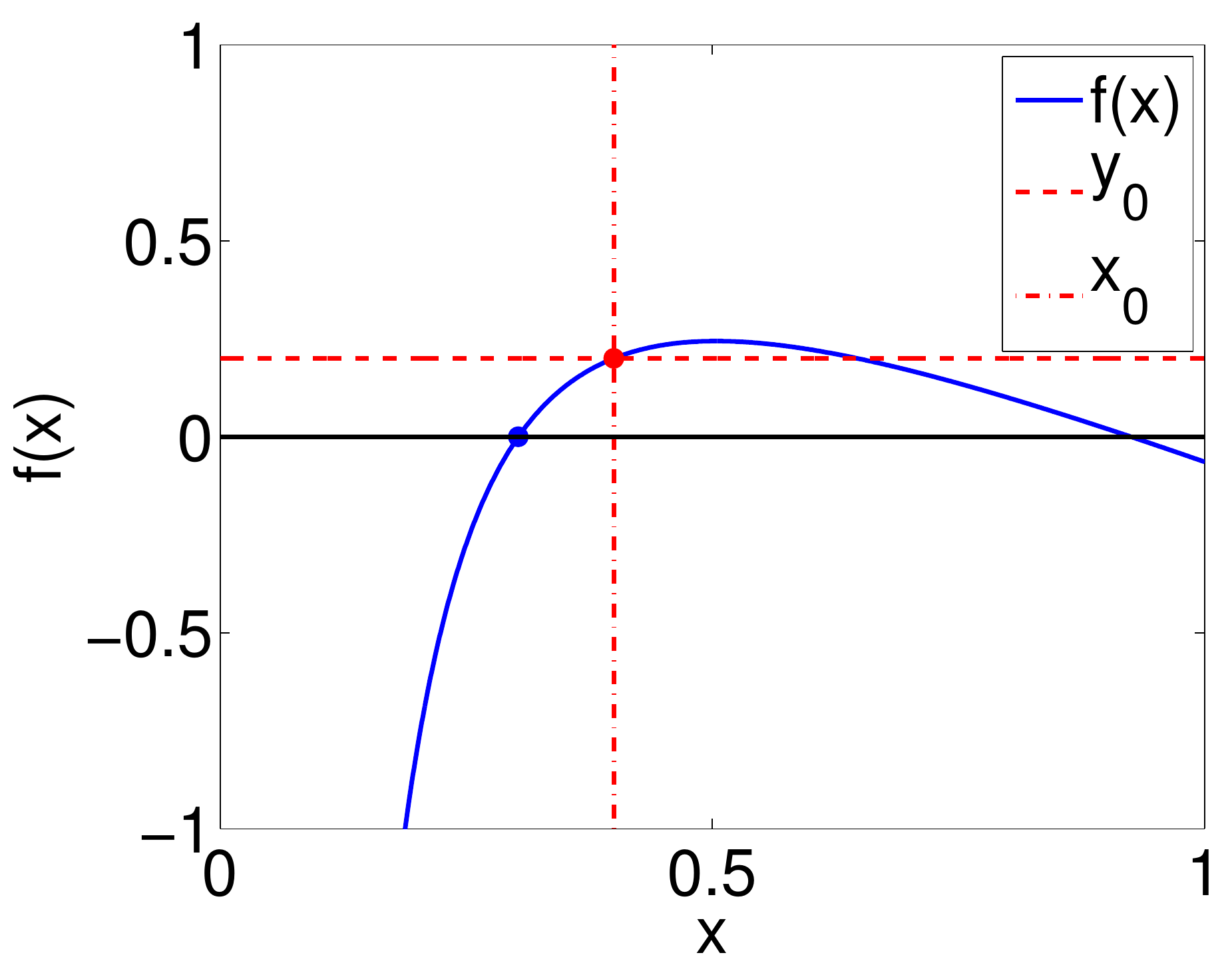}}
\subfigure[][$\rho < \frac{x_0}{z_0}$ and $y_0 = 0$.]{\includegraphics[width=0.23\linewidth]{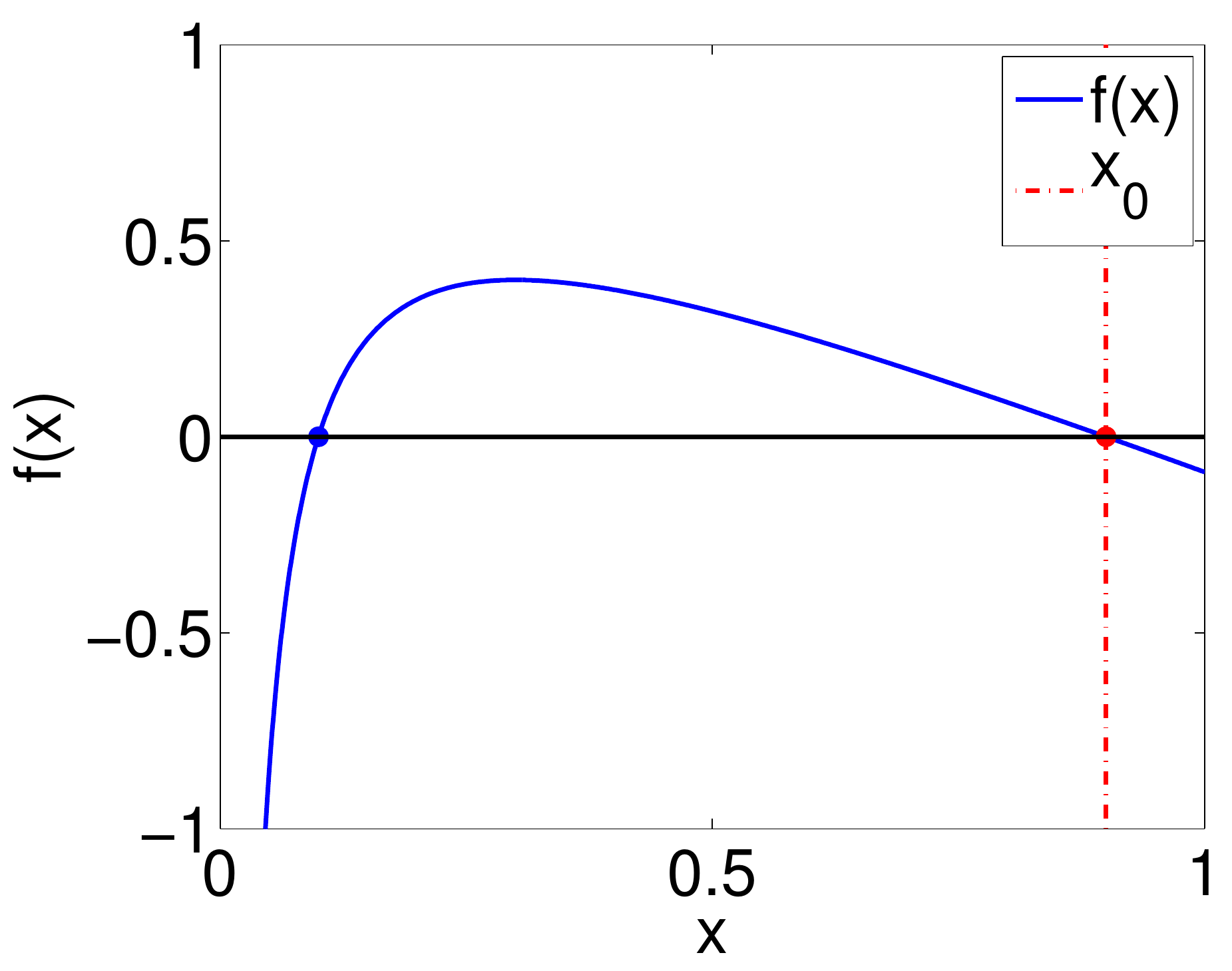}}
\subfigure[][$\rho > \frac{x_0}{z_0}$ and $y_0 = 0$.]{\includegraphics[width=0.23\linewidth]{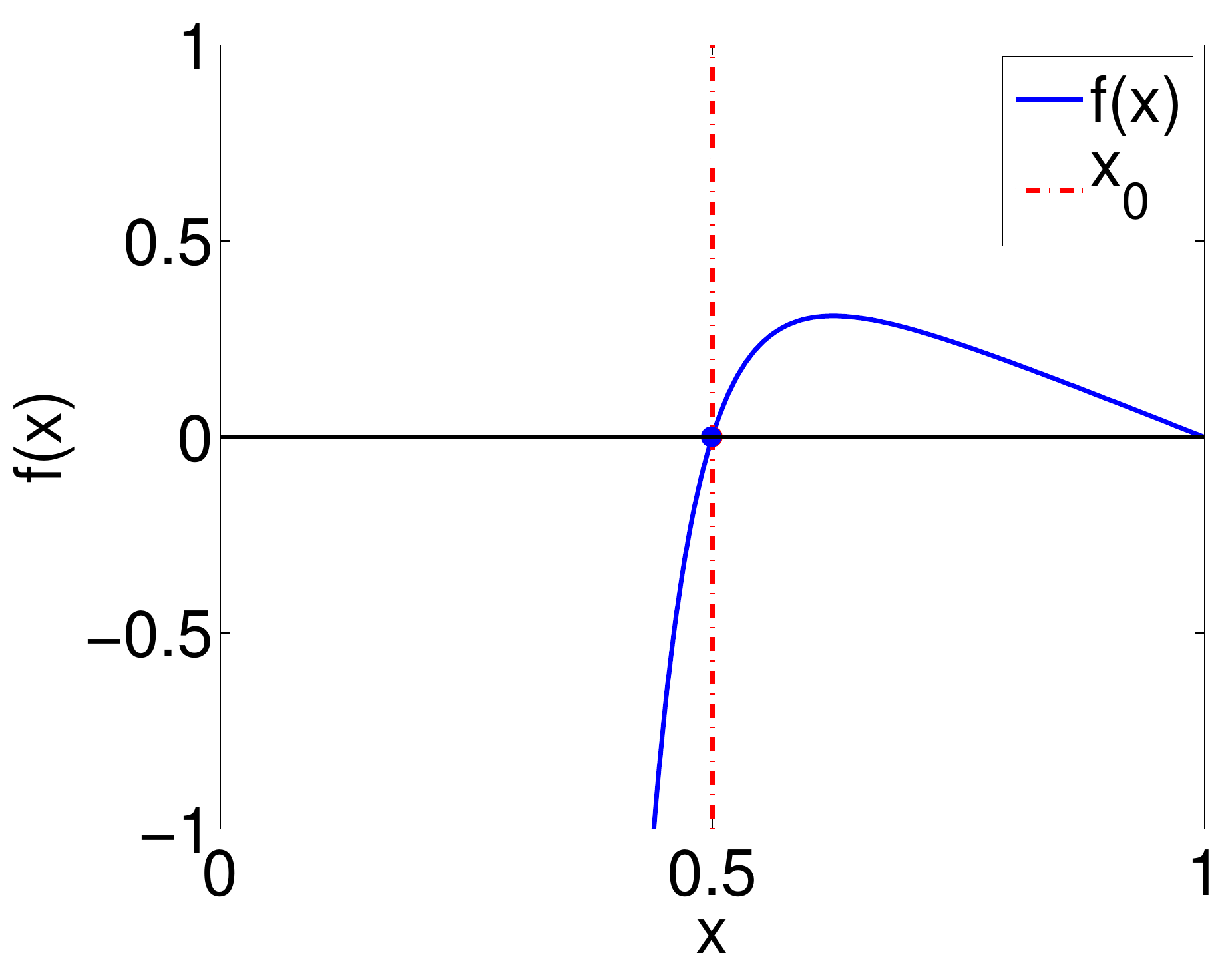}}
\end{center}
\caption{Four different cases for the function $f(x)$ given by Eq.~\eqref{eq:fx}.}
\label{fig:fx}
\end{figure*}

The first step in this direction is to define 
\begin{equation*}
 \theta^n(t) = \int_0^t Y^n(s) ds,
\end{equation*}
for $0 \leq t \leq \tau^{(n)}$. Notice that $ \theta^n$ is a strictly increasing, continuous and piecewise linear function. In this way, we can define its inverse by 
\begin{equation}
\Gamma^n(s) = \inf\{t: \theta^n(t) > s\}, 
\end{equation}
for $0 \leq s \leq \int_0^{\infty} Y^n(u) du$. Then it is not difficult to see that the process defined as
\begin{equation}
\tilde{V}^n(t) := V^n(\Gamma^n(t))
\end{equation}
has the same transitions as $\{ V^n(t) \}_{t \geq 0}$. As a consequence, if we define $\tilde{\tau}^n = \inf\{t: \tilde{Y}^n(t) = 0\}$ we get that $V^n(\tau^n) = \tilde{V}^n(\tilde{\tau}^n)$. This implies that it is enough to study $\tilde{X}^n (\tilde{\tau}^{(n)})/n$. The gain of the previous comparison relies on the fact that $\{ \tilde{V}^n(t) \}_{t \geq 0}$ is a continuous-time Markov chain with initial state $\left( x_0^n n, y_0^n n \right)$ and transitions and rates given by
\begin{equation*}
{\allowdisplaybreaks
\begin{array}{cc}
\text{transition} \quad &\text{rate} \\[0.1cm]
\ell_0 = (-1, 1) \quad &\lambda  X, \\[0.1cm]
\ell_1 = (0, -1) \quad &\alpha (n - X - Y).
\end{array}}
\end{equation*}
In particular, the rates of the process can be written as
\begin{equation*}
n \left[ \beta_{l_i} \left( \frac{ \tilde{X}}{n}, \frac{\tilde{Y}}{n} \right) \right],
\end{equation*}
where $\beta_0(x,y) = \lambda x$ and $\beta_1(x,y) = \alpha \left( 1 - x - y \right)$. Processes defined as above are called \emph{density dependent} since the rates depend on the density of the process (i.e. normed by $n$). Then $\{\tilde{V}^n(t)\}_{t\geq 0}$ is a density dependent Markov chain with possible transitions in the set $\{\ell_0,\ell_1\}$. By applying convergence results of~\cite{Ethier2009}, we obtain an approximation of this process, as the population size becomes larger, by a system of differential equations. More precisely, it is known that the limit behavior of the density dependent Markov chain $\{\tilde{V}^n(t)\}_{t\geq 0}$ can be determined by the drift function $F(x,y) = l_0 \beta_0 (x, y) + l_1 \beta_1 (x, y)$ (see the Appendix for more details). In other words,
\begin{equation}
F(x,y) = \left( -\lambda x, (\lambda + \alpha) x + \alpha y - \alpha \right)
\end{equation}
and the limiting system of ordinary differential equations is given by
\begin{equation}\label{Eq:edo}
\begin{cases}
x^{\prime}(t) = - \lambda x(t), \\[0.1cm]
y^{\prime}(t) = (\lambda + \alpha) x(t) + \alpha y(t) - \alpha, \\[0.1cm]
x(0) = x_0, y(0) = y_0.
\end{cases}
\end{equation}
The solution of~\eqref{Eq:edo} is
\begin{equation}
\begin{cases}
x(t) = x_0 \exp(-\lambda t), \\[0.1cm]
y(t) = f(x(t)),
\label{eq:xinf}
\end{cases}
\end{equation}
where $f:(0,x_0]\rightarrow \mathbb{R}$ is given by
\begin{equation}
f(x)=1 - (1 - x_0 - y_0) \left( \frac{x_0}{x} \right)^{\rho} - x.
\label{eq:fx}
\end{equation}
Figure~\ref{fig:fx} shows the behavior of $f(x)$ for four possible relations between $\rho$ and the initial conditions.
If $x_{\infty}$ denotes the root of $f(x) = 0$ in $(0, x_0]$, then
\begin{equation}\label{conv:probability}
\lim_{n\to \infty}\frac{X^n (\tau^n)}{n} =x_{\infty}
\end{equation}
in probability (see Appendix). This means that, for $n$ large enough, with high probability the process dies out leaving approximately a proportion $x_{\infty}$ of remaining ignorant nodes of the population. Furthermore, we can describe the distribution of the random fluctuations around the limiting value $x_{\infty}$. More precisely, by assuming that $y_0 > 0$, or that $y_0 = 0$ and $\rho < x_0 / z_0$, we obtain the following central limit theorem (see Appendix)
\begin{equation}
\label{CLT}
\sqrt{n} \left[\frac{X^n(\tau^{(n)})}{n} - x_\infty \right] \Rightarrow \mathcal{N}(0, \sigma^2) \, \text{ as } \, n \to \infty,
\end{equation}
where $ \Rightarrow $ denotes convergence in distribution and $ \mathcal{N}(0, \sigma^2) $ is the Gaussian distribution with mean zero and variance $\sigma^2:=\sigma^2(\alpha,\lambda,x_0,y_0,z_0)$ given by
\begin{equation}\label{variancia}
\frac{x_{\infty} z_{\infty} \left[x_0 \, x_{\infty} (1 - z_0 - x_{\infty}) + z_0 \, \rho^2 z_{\infty} (x_0 - x_{\infty})\right]}{x_0 \, z_0 \left[\rho -
x_{\infty} (\rho + 1)\right]^2},
\end{equation}
where $z_{\infty}:=1 - x_{\infty}$.

As mentioned previously, Kortchemski~\cite{kortchemski/2014} deals with this model on the complete graph in the context of epidemic spreading. More precisely, the case $X(0)=n$ and $Y(0)=Z(0)=1$ is considered in a population of size $n+2$. Interesting results related to limit theorems and phase transitions are obtained. The results stated here concerning the asymptotic behavior of the rumor process are proved under a different initial configuration and have a different convergence scale. We observe that the case considered in~\cite{kortchemski/2014} is, using our notation, $x_0 =1$ and $y_0=z_0=0$ (see equation \eqref{limites}). Therefore, our work complement the results by Kortchemski~\cite{kortchemski/2014}.

\section{Heterogeneously mixing populations}

As an interacting particle system, our model can be formulated in a finite graph (or network) $G$ as
a continuous-time Markov process $(\eta_t)_{t\geq 0}$ on the state space $\{0,1,2\}^V$, where $V:=\{1,2,\ldots, n\}$ is the set of nodes. A state of the process is a vector $\eta = (\eta(i) : i \in V)$, where $\eta(i)\in\{0,1,2\}$ and $0$, $1$, $2$ represent the ignorant, spreader and stifler states, respectively. The rumor is spread at rate $\lambda$ and a spreader becomes a stifler at rate $\alpha$ after contacting stiflers. We assume that the state of the process at time $t$ is $\eta$ and let $i\in V$. Then 
$$P(\eta_{t+h}(i)=1|\eta_t(i) =0)=\lambda h N_1(i) + o(h)$$
$$P(\eta_{t+h}(i)=2|\eta_t(i) =1)=\alpha h N_2(i) + o(h)$$
where $N_{\ell}(i):=N_{\ell}(\eta,i)$ is the number of neighbors of $i$ that are in state $\ell$, for $\ell=1,2$ and for the configuration $\eta$. In the previous section we present a rigorous analysis of our rumor model on a complete graph with $n$ vertices. Our results in such case are related to the asymptotic behavior of the random variables
\begin{align*}
X^{(n)}(t) &=\sum_{i=1}^{n}I_{\{\eta_t(i)=0\}},\\
Y^{(n)}(t) &=\sum_{i=1}^{n}I_{\{\eta_t(i)=1\}},
\end{align*}
where $I_{A}$ denotes the indicator random variable of the event $A$.
This mean-field approximation assumes that the possible contacts between each pair of individuals occur with the same probability. This assumption enables an analytical treatment, but does not represent the organization of real-world networks, whose topology is very heterogeneous~\cite{Boccaletti06:PR, Costa011:AP}. In this case, we use a different approach that allows us to describe the evolution of each node. Such formulation assumes the independence among the state of the nodes. More precisely, we are interested in the behavior of the probabilities
\begin{equation} \label{probabilities}
\begin{split}
x_i(t) &:=P(\eta_t(i)=0),\\
y_i(t) &:=P(\eta_t(i)=1),\\
z_i(t) &:=P(\eta_t(i)=2),
\end{split}
\end{equation}
for all $i=1,2,\ldots,n$. We describe our process in terms of a collection of independent Poisson processes $N_i^{\lambda}$ and $N_i^{\alpha}$ with intensities $\lambda$ and $\alpha$, respectively, for $i=1,2,\ldots,n.$  We associate the processes $N_i^{\lambda}$ and $N_i^{\alpha}$ to the node $i$ and we say that at each time of $N_i^{\lambda}$ ($N_i^{\alpha}$), if $i$ is in state $1$ ($2$) then it choses a nearest neighbor $j$ at random and tries to transmit (scotch) the information provided $j$ is in state $0$ (1). In this way, we obtain a realization of our process $(\eta_t)_{t\geq 0}$.

In order to study the evolution of the functions \eqref{probabilities}, we fix a node $i$ and  analyze the behavior of its different transition probabilities on a small time window. More precisely, consider a small enough positive number $h$ and note that
\begin{equation}\label{Eq:probcond1}
P(\eta_{t+h}(i)=0)=P(\eta_{t+h}(i)=0|\eta_{t}(i)=0)P(\eta_{t}(i)=0),
\end{equation}
where the first factor of the right-hand side of last expression is given by
\begin{widetext}
\begin{equation}\label{Eq:probcond2}
\begin{array}{rcl}
P(\eta_{t+h}(i)=0|\eta_{t}(i)=0) &=&1-P(\eta_{t+h}(i)=1|\eta_{t}(i)=0)-P(\eta_{t+h}(i)=2|\eta_{t}(i)=0)\\[.2cm]
& =&1-P(\eta_{t+h}(i)=1|\eta_{t}(i)=0) + o(h).
\end{array}
\end{equation}
\end{widetext}

The $o(h)$ term appears in the above equation, because the occurrence of a transition from state $0$ to state $2$ in a time interval of size $h$ implies the existence of at least two marks of a Poisson process at the same time interval. On the other hand, if we denote $\mathcal{B}_{ji}(h)$ as the intersection of the events $\{N_{j}(t,t+h)=1\}$, $\{j \text{ transmit the information to }i \text{ in }(t,t+h)\}$, $\{\eta_j(t)=1\}$ and $\{\eta_j(s)=1, \text{ for }t<s\leq t+h\}$, we obtain
\begin{widetext}
\begin{equation}\label{Eq:probcond3}
\begin{array}{rcl}
P(\eta_{t+h}(i)=1|\eta_{t}(i)=0) &= & P\left(\eta_{t+h}(i)=1|\eta_{t}(i)=0,\cup_{j=1}^{n}\mathcal{B}_{ji}(h)\right)P(\mathcal{B}_{ji}(h)|\eta_{t}(i)=0) + o(h),\\[.2cm]
&=&\sum_{j=1}^{n}\frac{A_{ji}}{k_j}(\lambda h+o(h))P(\eta_t(j)=1) + o(h),
\end{array}
\end{equation}
\end{widetext}
where $A_{ji}=1$ if $i$ is a direct neighbor of  $j$ in the network (equals $0$ other case) and $k_i = \sum_j A_{ij}$ is the degree of the node $i$. Thus, we obtain
\begin{widetext}
$$P(\eta_{t+h}(i)=0)=\left(1-\sum_{j=1}^{n}\frac{A_{ji}}{k_j}\left(\lambda h+o(h)\right)P(\eta_t(j)=1) + o(h)\right)P(\eta_{t}(i)=0)$$
or
$$P(\eta_{t+h}(i)=0)-P(\eta_{t}(i)=0)=-\left(\sum_{j=1}^{n}\frac{A_{ji}}{k_j}(\lambda h+o(h))P(\eta_t(j)=1) + o(h)\right)P(\eta_{t}(i)=0).$$
\end{widetext}
Finally, as $x_i^{\prime}(t)=\lim_{h\to 0}(x_i(t+h)-x_i(t))/h$ we conclude $x_i^{\prime}(t)=-\lambda x_i(t) \sum_{j=1}^{n} \frac{A_{ji}}{k_j} y_j(t).$ Same arguments allow us to obtain the equations for $y_i(t)$ and $z_i(t)$. In this way, we have the following set of dynamical equations
\begin{equation}
\begin{cases}
x_i^{\prime}(t) = - \lambda x_i(t) \sum_{j=1}^n P_{ji} y_j(t), \\[0.1cm]
y_i^{\prime}(t) = \lambda x_i(t) \sum_{j=1}^n P_{ji} y_j(t) - \alpha y_i(t) \sum_{j=1}^n P_{ji}  z_j(t), \\[0.1cm]
z_i^{\prime}(t) = \alpha y_i(t) \sum_{j=1}^n P_{ji}  z_j(t), \\[0.1cm]
x_i(0) = x_0, y_i(0) = y_0, z_i(0) = z_0,
\end{cases}
\label{eq:network_edo}
\end{equation}
for all $i=1,2,\ldots,n$, and $P_{ji}:=A_{ji}/k_j.$ We observe that when the network considered is a complete graph of $n$ vertices, the system of equations~\eqref{eq:network_edo} match with the homogeneous approach (see the system of equations~\eqref{eq:edo}).

\begin{figure*}[!t]
\begin{center}
\subfigure[][]{\includegraphics[width=0.32\linewidth]{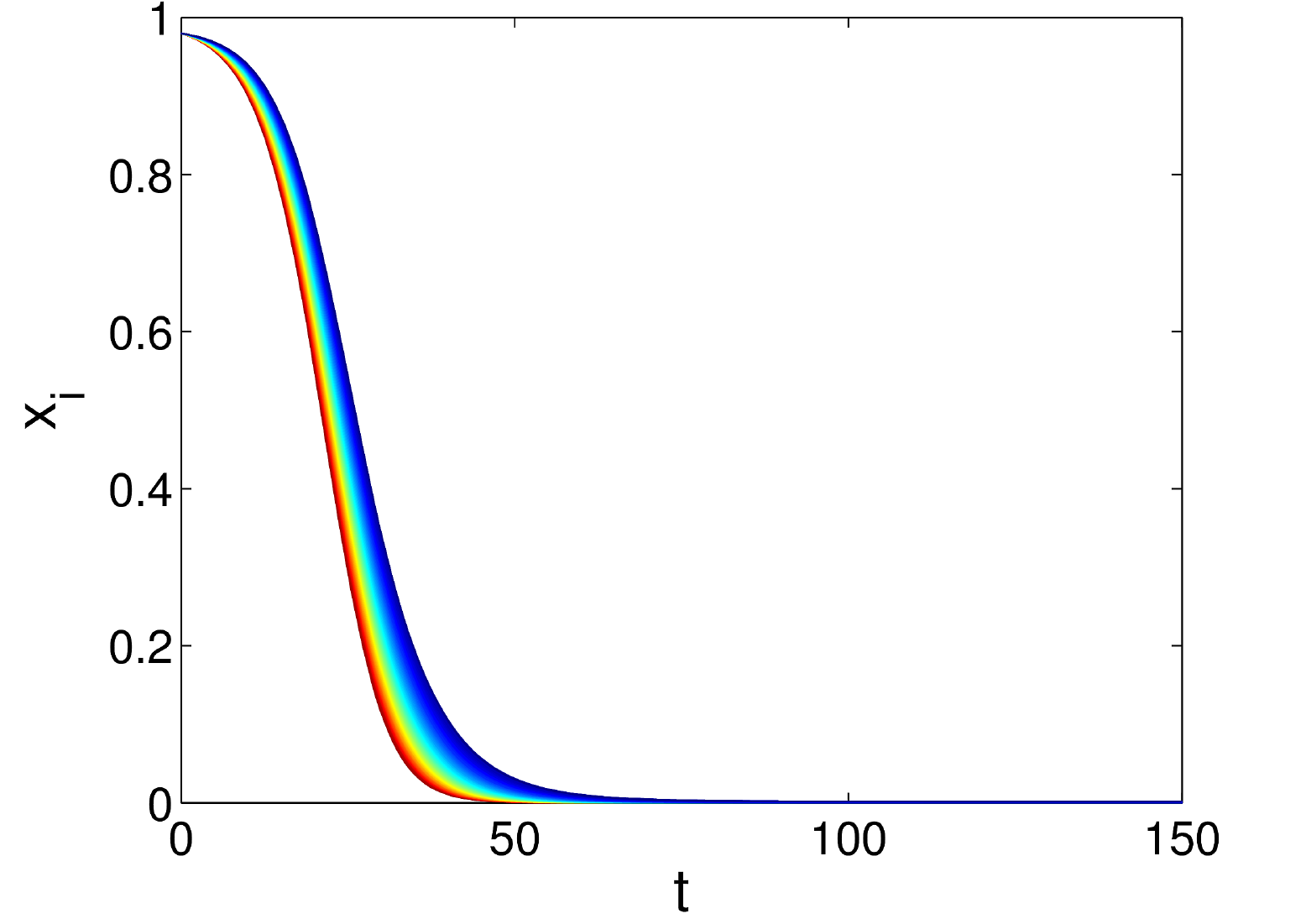}}
\subfigure[][]{\includegraphics[width=0.32\linewidth]{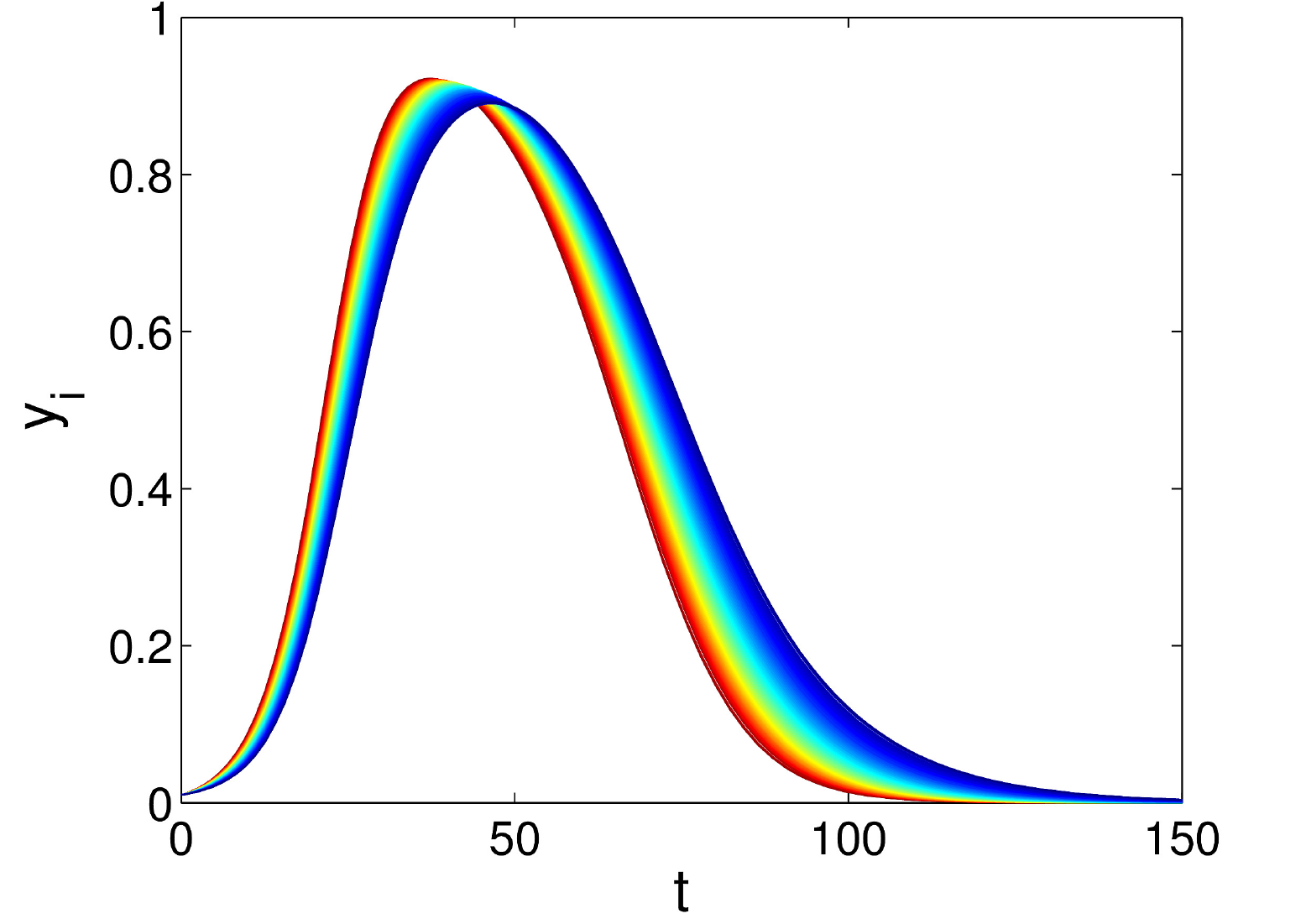}}
\subfigure[][]{\includegraphics[width=0.32\linewidth]{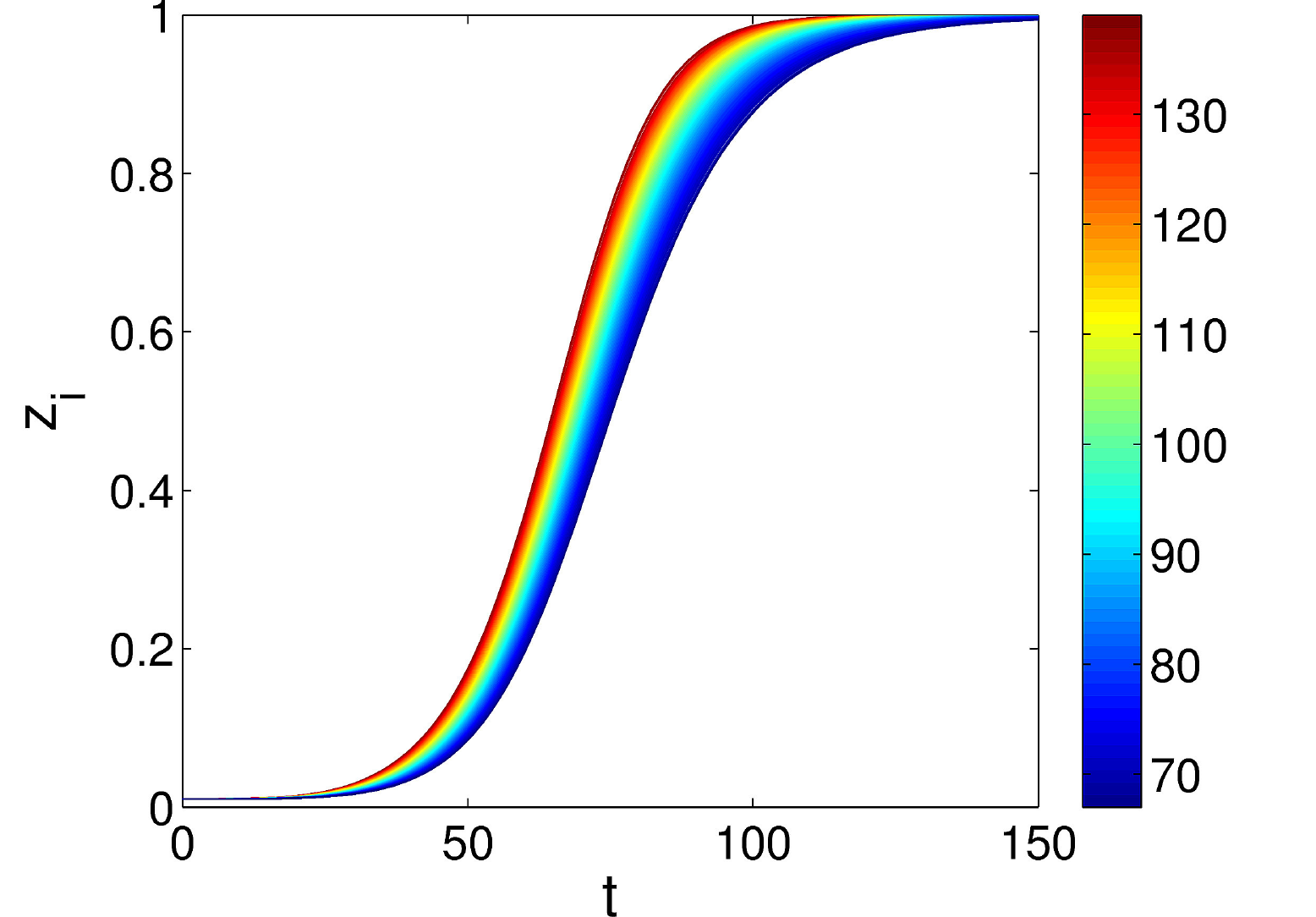}}
\end{center}
\caption{Time evolution of the nodal probabilities considering our model for an Erd\H{o}s and R\'{e}nyi network with $n = 10^4$ nodes and $\langle k \rangle \approx 100$. We consider the spreading rate $\lambda = 0.2$ and stifling rate $\alpha = 0.1$. Each curve represents the probability that a node is in one of the three states (ignorant, spreader or stifler) and the color represents the degree of the node $i$. The initial conditions are $x_0 = 0.98$, $y_0 = 0.01$ and $z_0 = 0.01$.}
\label{fig:ER}
\end{figure*}

\begin{figure*}[!t]
\begin{center}
\subfigure[][]{\includegraphics[width=0.32\linewidth]{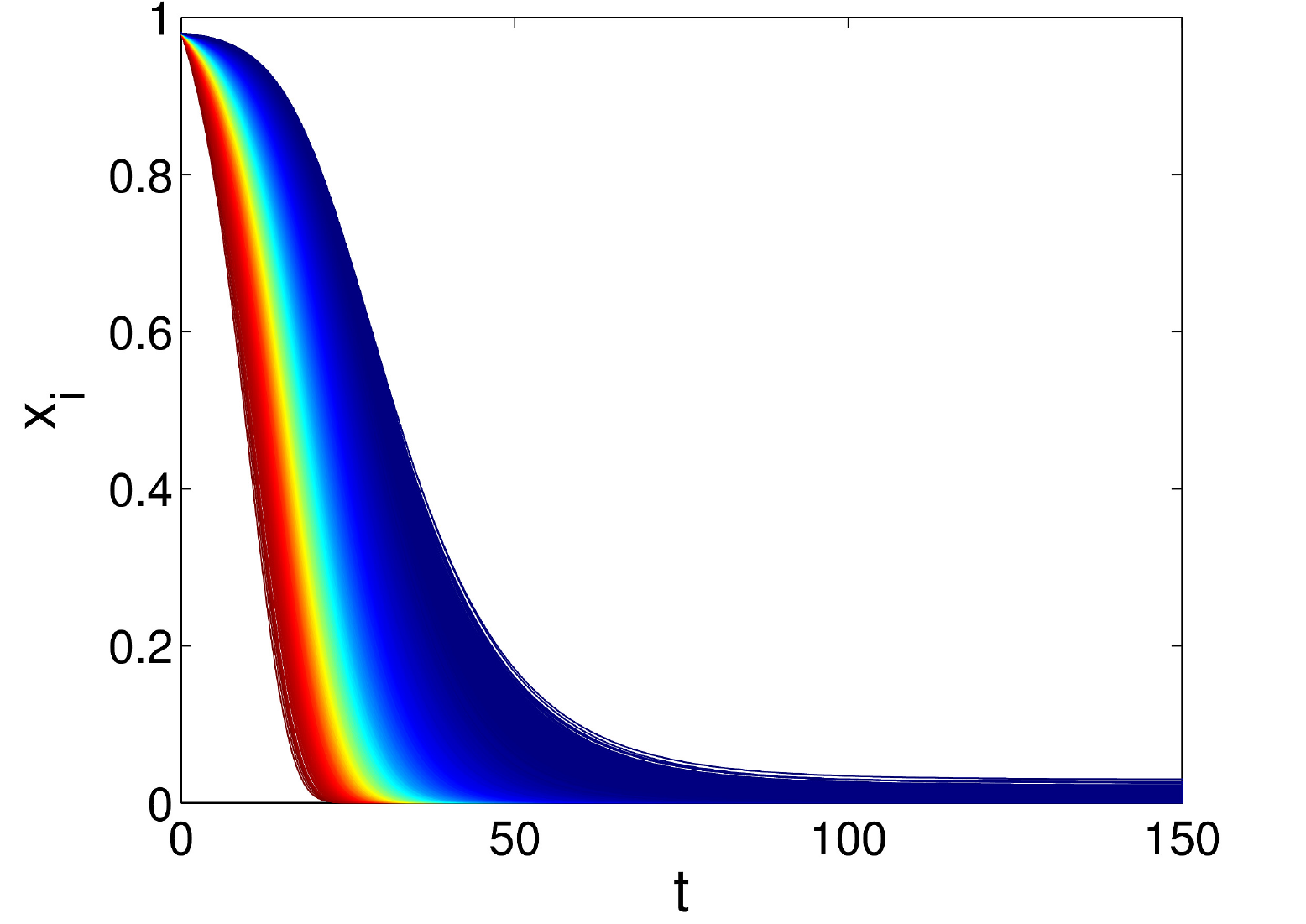}}
\subfigure[][]{\includegraphics[width=0.32\linewidth]{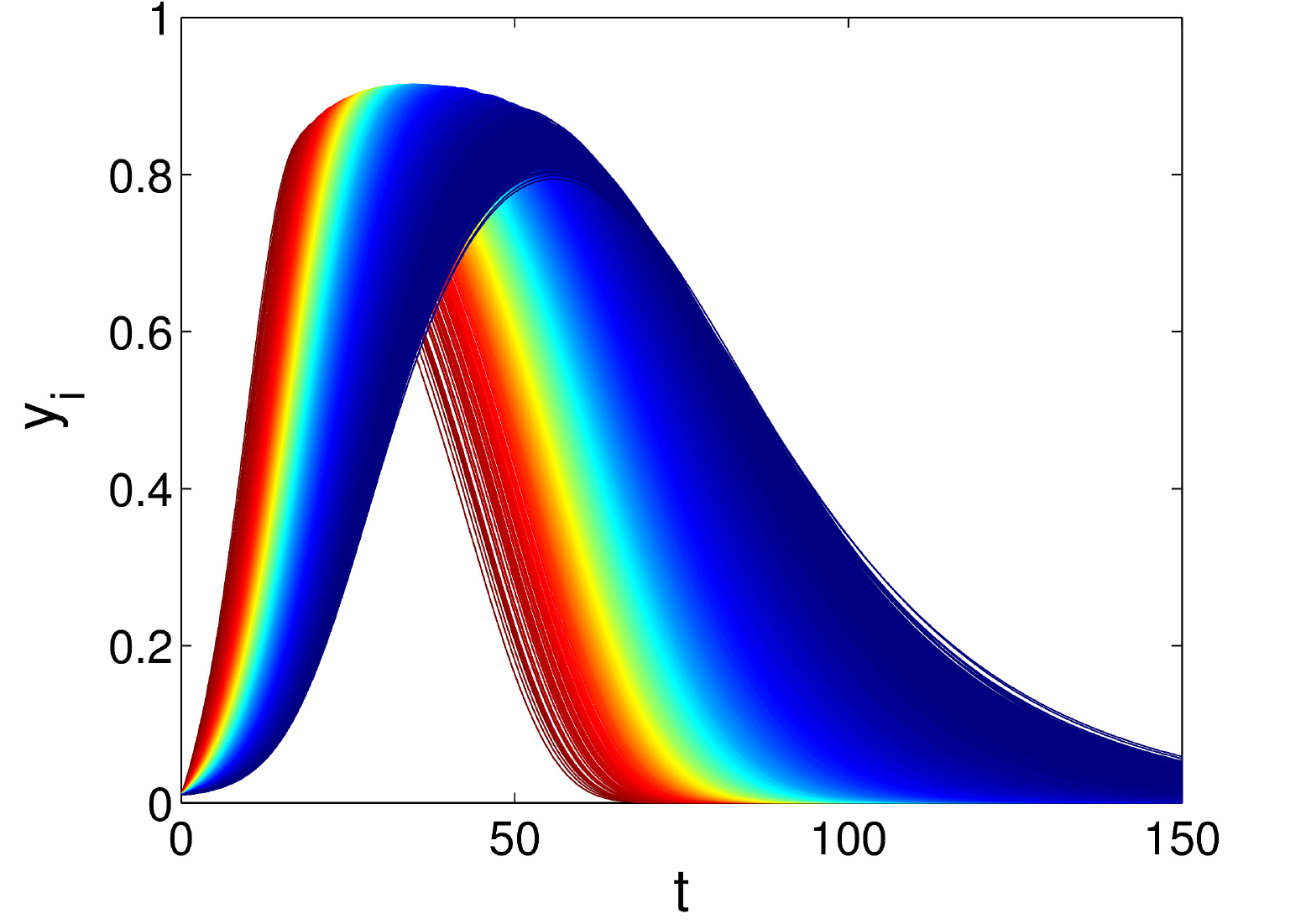}}
\subfigure[][]{\includegraphics[width=0.32\linewidth]{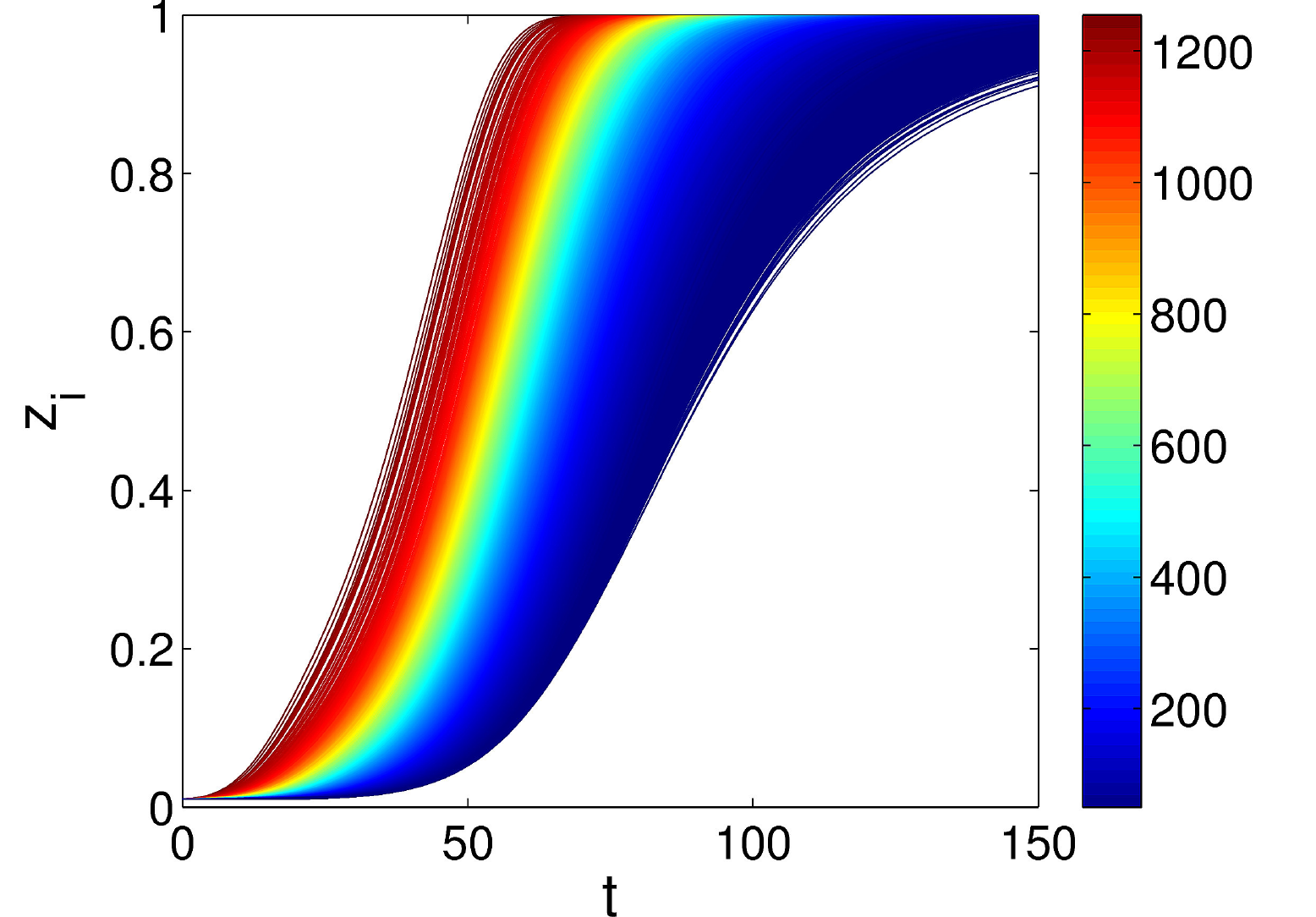}}
\end{center}
\caption{Time evolution of the nodal probabilities considering our model for an Barab\'{a}si and Albert network with $n = 10^4$ nodes and $\langle k \rangle \approx 100$. The spreading rate as $\lambda = 0.2$, while the stifling rate is $\alpha = 0.1$. Each curve represents the probability that a node is in one of the three states (ignorant, spreader or stifler) and the color represents the degree of the node $i$. The initial conditions are $x_0 = 0.98$, $y_0 = 0.01$ and $z_0 = 0.01$.}
\label{fig:BA}
\end{figure*}

In order to verify the influence of network structure on the dynamical behavior of the models, we consider random graphs of Erd\H{o}s and R\'{e}nyi (ER) and scale-free networks of Barabási and Albert (BA). Random graphs are created by a Bernoulli process, connecting each pair of vertices with the same probability $p$. The degree distribution of random graphs follows a Poisson distribution for large values of $n$ and small $p$, as a consequence of the law of rare events~\cite{erdos1959random}. On the other hand, the BA model generates scale-free networks by taking into account the network growth and preferential attachment rules~\cite{Barabasi99}. The networks generated by this model present degree distribution following a power-law, $P(k) \sim k^{-\gamma}$, with $\gamma  = 3$. In random graphs most of the nodes have similar degrees, whereas scale-free networks are characterized by a very heterogeneous  structure.

Figures~\ref{fig:ER} and~\ref{fig:BA} show the time evolution of the nodal probabilities, considering ER and BA networks, respectively. These results are obtained by solving numerically the system of equations~\eqref{eq:network_edo}. Both networks have $n = 10^4$ nodes and $\langle k \rangle \approx 100$. The spreading rate is $\lambda = 0.2$ and the stifling rate is $\alpha = 0.1$. The color of each curve denotes the degree of each node $i$. Comparing Figures~\ref{fig:ER} and~\ref{fig:BA}, we can see that the variance of $x_i, y_i$ and $z_i$ in BA networks is higher than in ER networks. Moreover, in both networks, higher degree nodes tend to turn into a stifler earlier than lower degree ones.

We compare the behavior of our model, described by Equation~\ref{eq:network_edo}, with the Maki and Thompson model~\cite{maki/thompson/1973} in ER and BA networks.  The time evolution of this model is given by 
\begin{equation}
\begin{cases}
x_i^{\prime}(t) = - \lambda x_i(t) \sum_{j=1}^n P_{ji} y_j(t), \\[0.1cm]
y_i^{\prime}(t) = \lambda x_i(t) \sum_{j=1}^n P_{ji} y_j(t) +  \\
\hspace{1.2cm} - \alpha y_i(t)\sum_{j=1}^n P_{ij}  \left( x_j(t) + z_j(t) \right), \\[0.1cm]
z_i^{\prime}(t) = \alpha y_i(t) \sum_{j=1}^n P_{ij}  \left( x_j(t) + z_j(t) \right), \\[0.1cm]
x_i(0) = x_0, y_i(0) = y_0, z_i(0) = z_0,
\end{cases}
\label{eq:network_edo_mt}
\end{equation}
where, as before, $x_i$, $y_i$ and $z_i$ are the micro-state variables, quantifying the probability that the node $i$ is an ignorant, spreader or a stifler at time $t$, respectively, for $i=1,2,\ldots, n$. Note $x_i(t) + y_i(t) + z_i(t) = 1, \forall i, t$. 

\begin{figure*}[!t]
\begin{center}
\subfigure[][]{\includegraphics[width=0.32\linewidth]{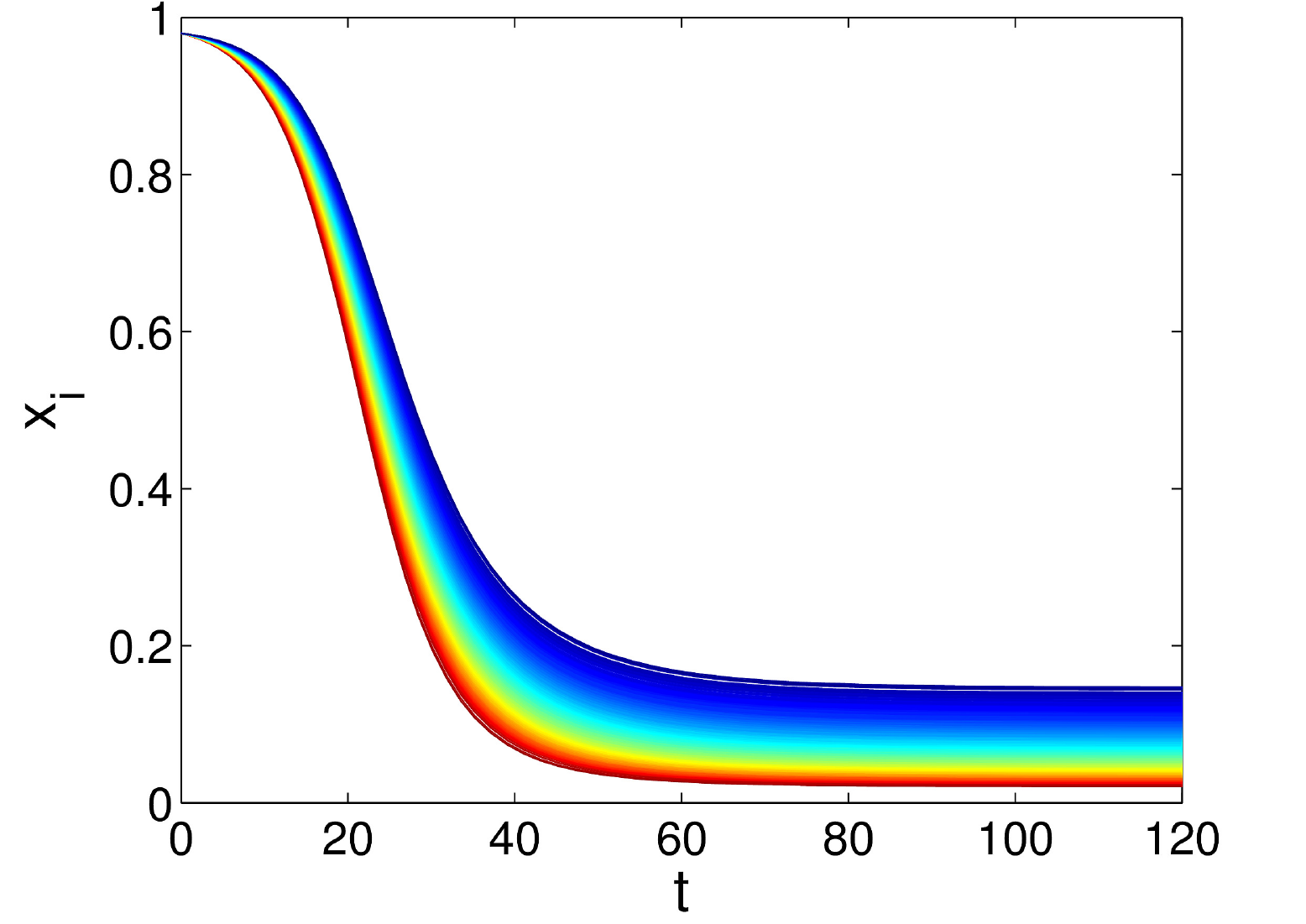}}
\subfigure[][]{\includegraphics[width=0.32\linewidth]{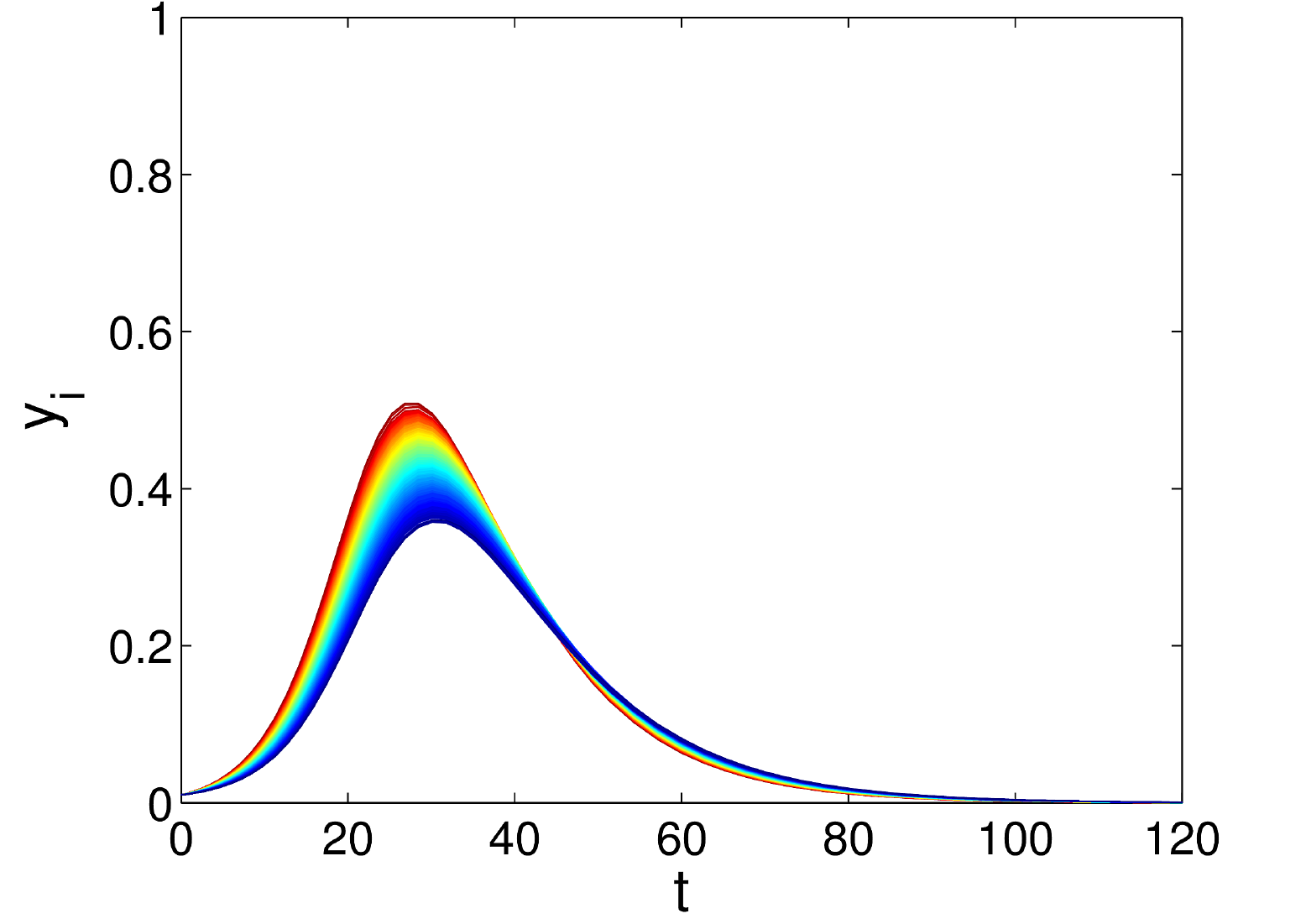}}
\subfigure[][]{\includegraphics[width=0.32\linewidth]{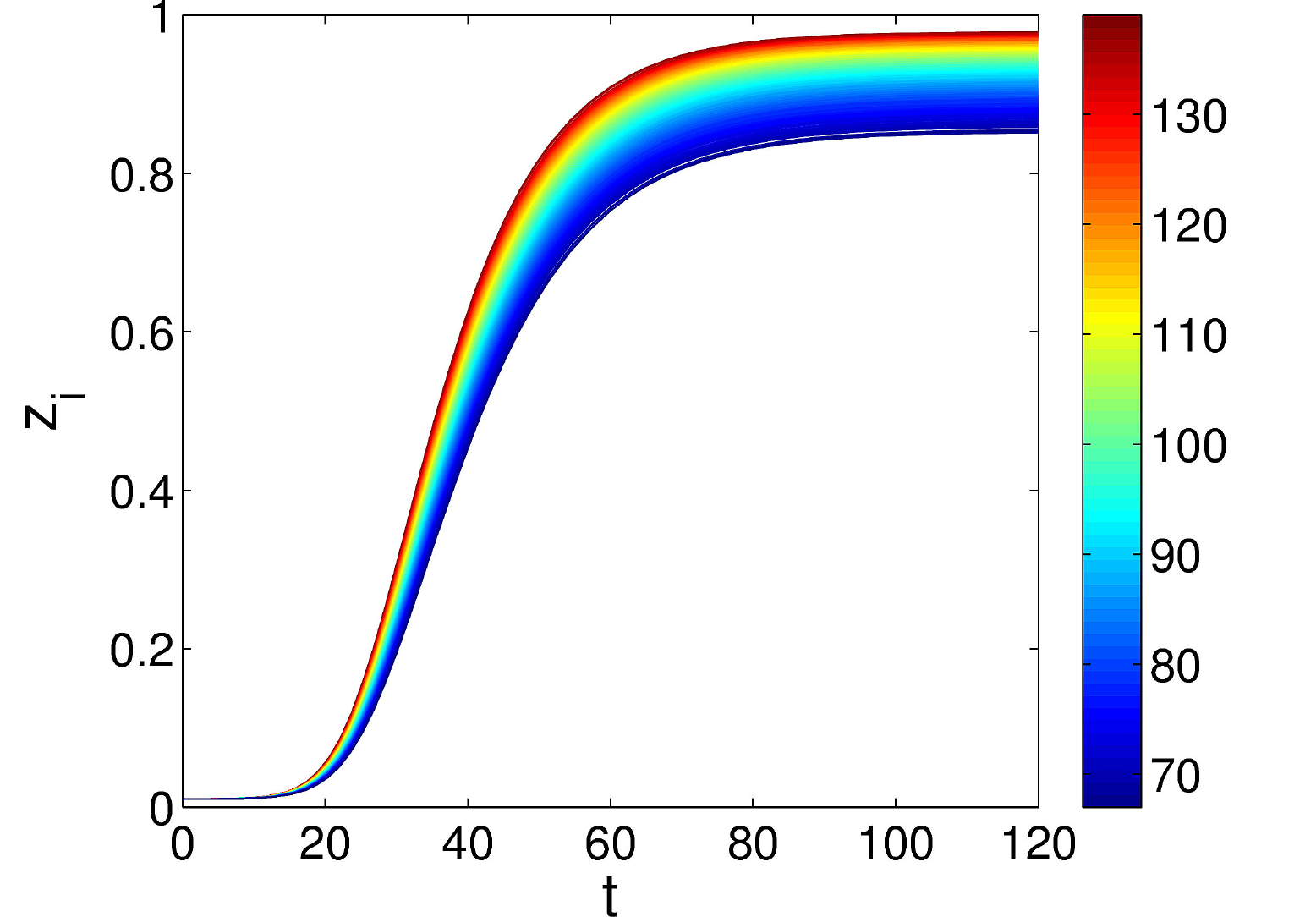}}
\end{center}
\caption{Time evolution of the nodal probabilities considering the Maki and Thompson model in an Erd\H{o}s and R\'{e}nyi network with $n = 10^4$ nodes and $\langle k \rangle \approx 100$. The spreading rate is $\lambda = 0.2$ and the stifling rate is $\alpha = 0.1$. Each curve represents the probability that a node is in one of the three states (ignorant, spreader or stifler) and the color represents the degree of the node $i$. The initial conditions are $x_0 = 0.98$, $y_0 = 0.01$ and $z_0 = 0.01$.}
\label{fig:ER_MT}
\end{figure*}

\begin{figure*}[!t]
\begin{center}
\subfigure[][]{\includegraphics[width=0.32\linewidth]{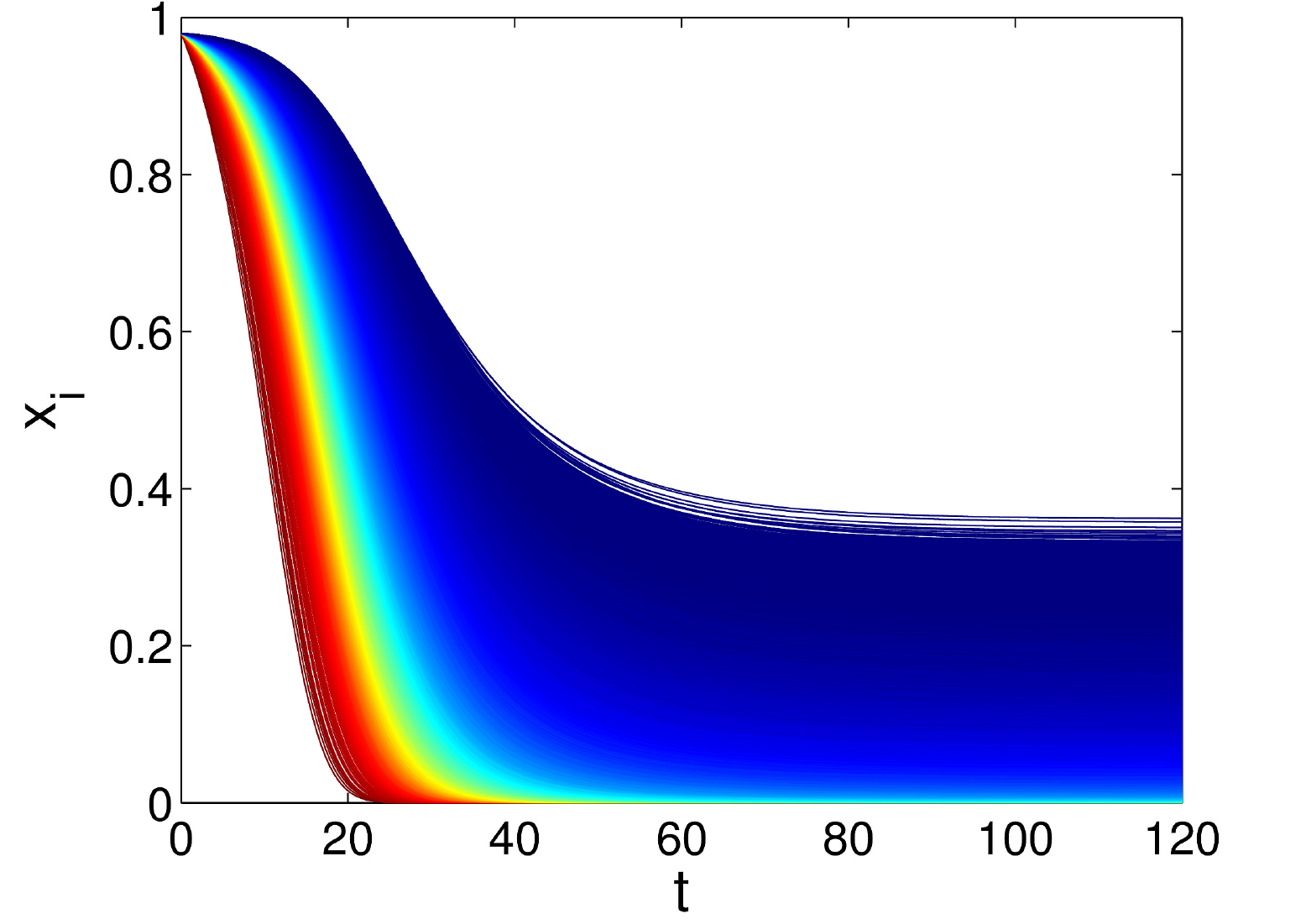}}
\subfigure[][]{\includegraphics[width=0.32\linewidth]{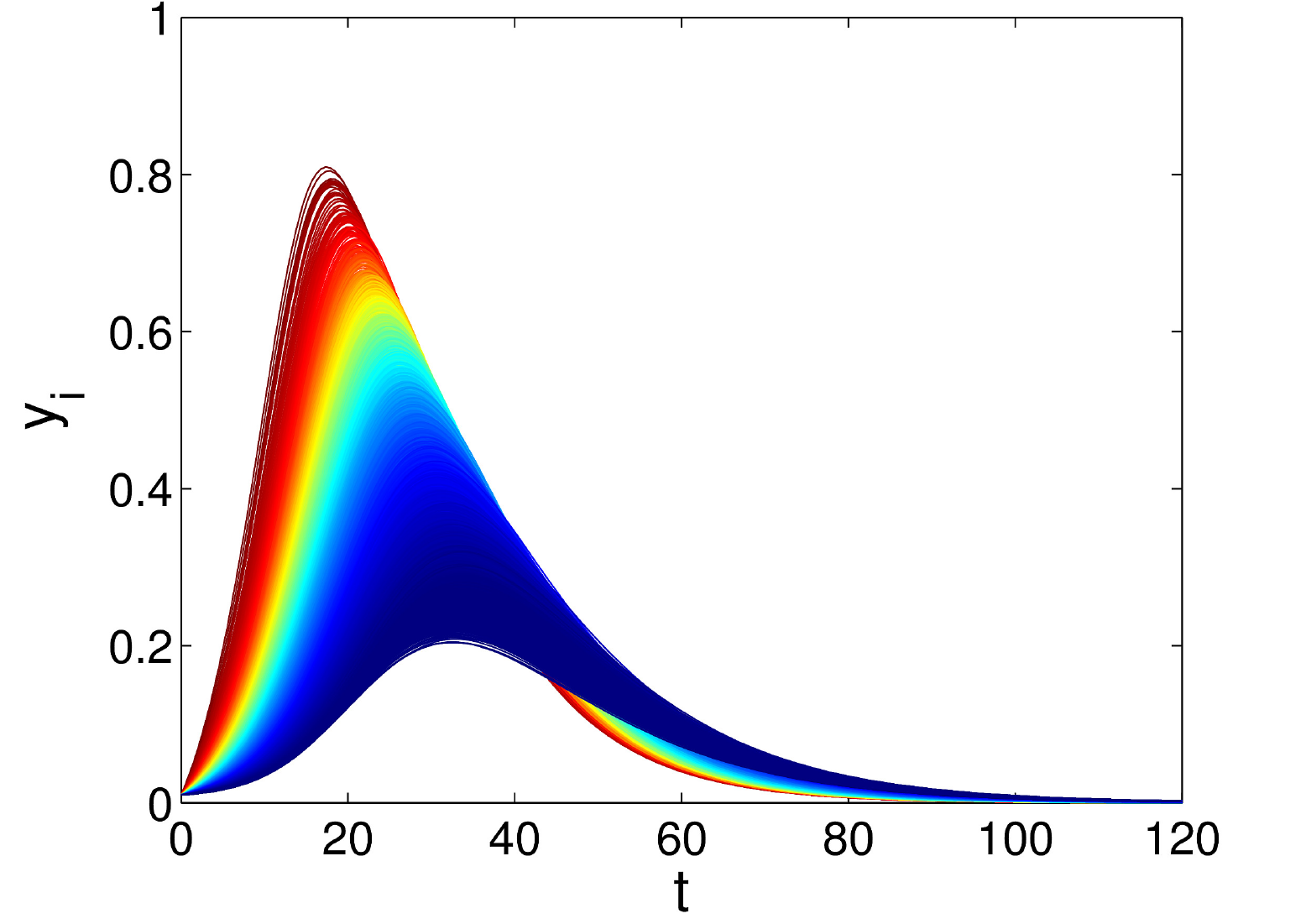}}
\subfigure[][]{\includegraphics[width=0.32\linewidth]{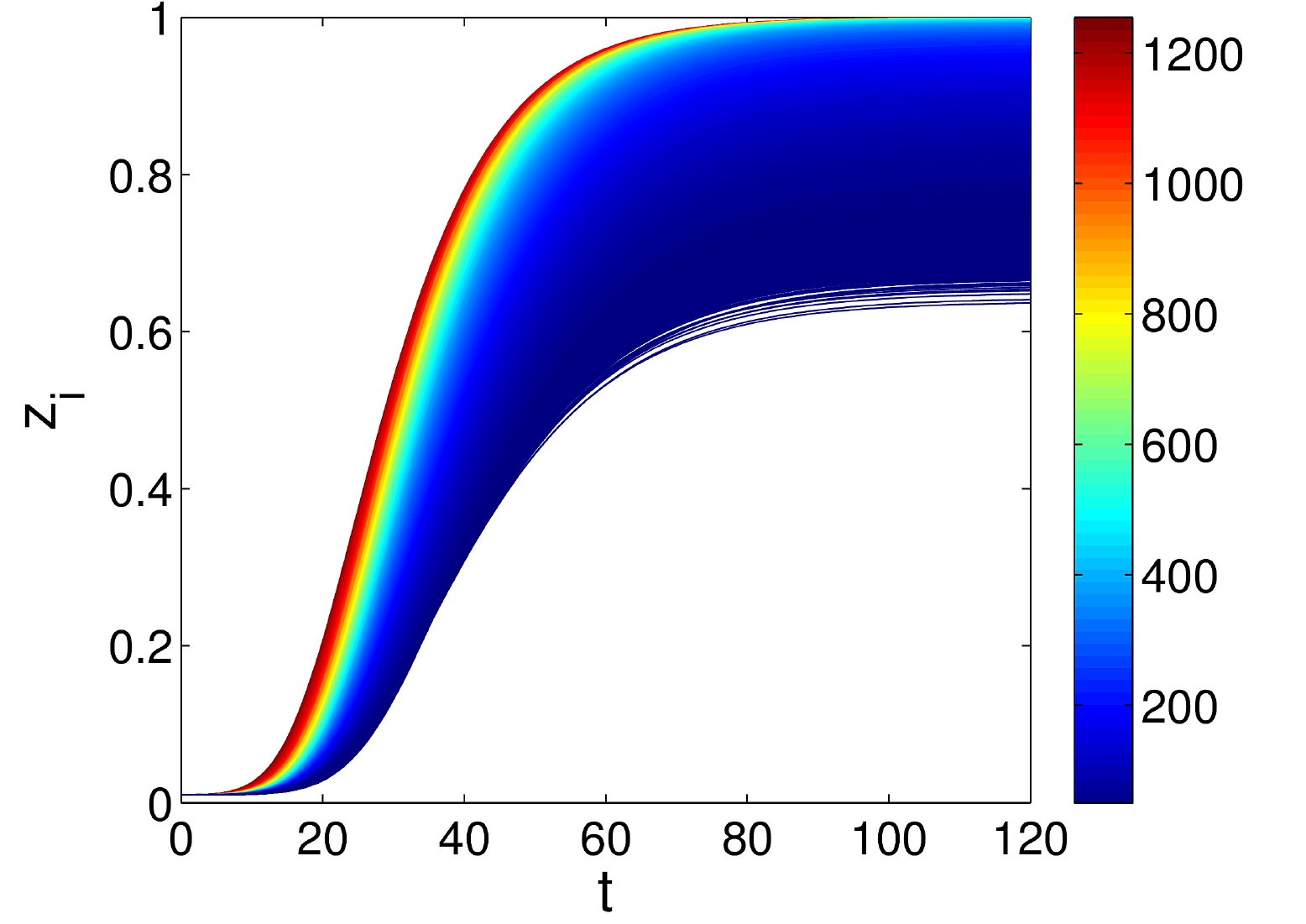}}
\end{center}
\caption{Time evolution of the nodal probabilities considering the Maki and Thompson model in an Barab\'{a}si and Albert network with $n = 10^4$ nodes and $\langle k \rangle \approx 100$. The spreading rate is $\lambda = 0.2$ and the stifling rate is $\alpha = 0.1$.  Each curve represents the probability that a node is in one of the three states (ignorant, spreader or stifler) and the color represents the degree of the node $i$. The initial conditions are $x_0 = 0.98$, $y_0 = 0.01$ and $z_0 = 0.01$.}
\label{fig:BA_MT}
\end{figure*}

Figures~\ref{fig:ER_MT} and~\ref{fig:BA_MT} show the time evolution of the nodal probabilities, by numerically solving equation~\eqref{eq:network_edo_mt}. Similarly to our model, the variances of in BA networks are higher than in ER networks. Besides, the hubs and leaves of the BA networks presents a completely different behavior, as can be seen in Figure~\ref{fig:BA_MT} (b). Moreover, the nodes having higher degrees also tend to become stifler earlier than low degree nodes.

We consider the same initial conditions for both rumor models, i.e. $x_0 = 0.98$, $y_0 = 0.01$ and $z_0 = 0.01$. It is worth emphasizing that the initial conditions in Figures~\ref{fig:ER_MT} and~\ref{fig:BA_MT} are not usual in the MT model, since most of the works on this model considers the initial fraction of stiflers as zero~\cite{Castellano09}. However, our model needs an initial non-zero fraction of stiflers, otherwise there is no manner to contain the rumor propagation. Furthermore, we can see that the peak of the fraction of spreaders in our model is higher than in the MT model. Such feature evinces the differences between two formulations. In the MT model the spreaders lose the interest in the rumor propagation due to the contact with individuals who have already known the rumor, whereas in our model spreaders are convinced only by stifler vertices to stop spreading the information. 

We can obtain a macro-sate variable to summarize the large-scale dynamical behavior of the system as the average over all states, i.e.,
\begin{equation}
 \phi^X = \frac{1}{n} \sum_{i=1}^n x_i,
\end{equation}
where $x_i$ is the probability that node $i$ is ignorant. Such quantity can be defined similarly for spreader, $y_i$, and stifler, $z_i$, nodes.

\section{Monte Carlo simulation}

Some results obtained from homogeneously mixing populations assumption can be extended to heterogeneous networks with relative accuracy on disassortative networks even when the mean degree is low~\cite{Gleeson012}. 
In this way, we perform extensive numerical simulations to verify how our rigorous results obtained for homogeneously mixing populations can be considered as approximations for random graphs and scale-free networks. The rumor spreading simulation is based on the contact between two individuals. At each time step each spreader makes a trial to spread the rumor to one of its neighbors and each stifler makes a trial to stop the spreading. If the spreader contacts an ignorant, it spreads the rumor with probability $\lambda$. Similarly, if the stifler contacts an spreader, that spreader becomes a stifler with probability  $\alpha$. The updates are performed in a sequential asynchronous fashion. For the simulation procedure it is important to randomize the state of the initial conditions, especially for the heterogeneous networks. In order to overcome statistical fluctuations in our simulations, every model is simulated 50 times with random initial conditions.

\subsection{Complete graph}

\begin{figure}[!t]
\begin{center}
\includegraphics[width=1\linewidth]{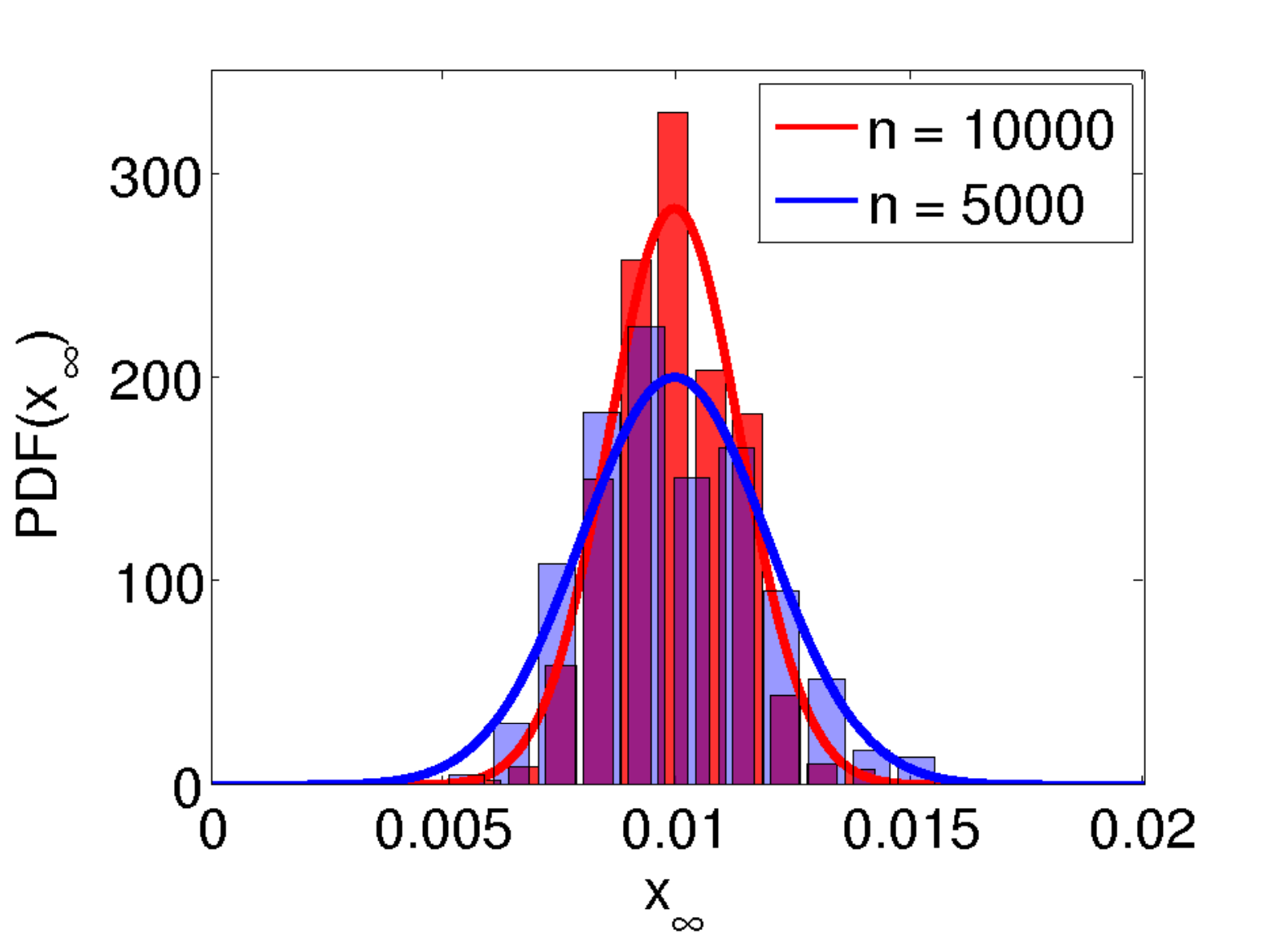}
\end{center}
\caption{Distribution of the fraction of ignorants obtained from 1000 simulations in a complete graph varying the number of nodes. The bars are obtained experimentally, while the fitted Gaussian are based on the theoretical values obtained from Eqs.~\eqref{eq:edo},~\eqref{eq:fx} and~\eqref{CLT}.}
\label{fig:tlc_full}
\end{figure}

The results are quantified as a function of the fraction of ignorant nodes, since when the time tends to infinity, the proportion of spreaders tends to zero and the fraction of ignorants and stiflers has complementary information about the population. Figure~\eqref{fig:tlc_full} compares the distribution of the fraction of ignorants obtained by Monte Carlo simulations with the central limit theorem by fitting a Gaussian distribution according to the theoretical values obtained from Eqs.~\eqref{eq:edo},~\eqref{eq:fx} and~\eqref{CLT}. Complete graphs of two different sizes are considered to show the dependency on the number of nodes $n$. Note that Eqs.~\eqref{eq:fx} and~\eqref{CLT} assert that only the variance depends on the network size, i.e. $\sigma^2 \propto \frac{1}{\sqrt{n}}$. Thus, the numerical simulations agree remarkably with the theoretical results.

\subsection{Complex networks}

\begin{figure*}[!t]
\begin{center}
\subfigure[]{\includegraphics[width=0.45\linewidth]{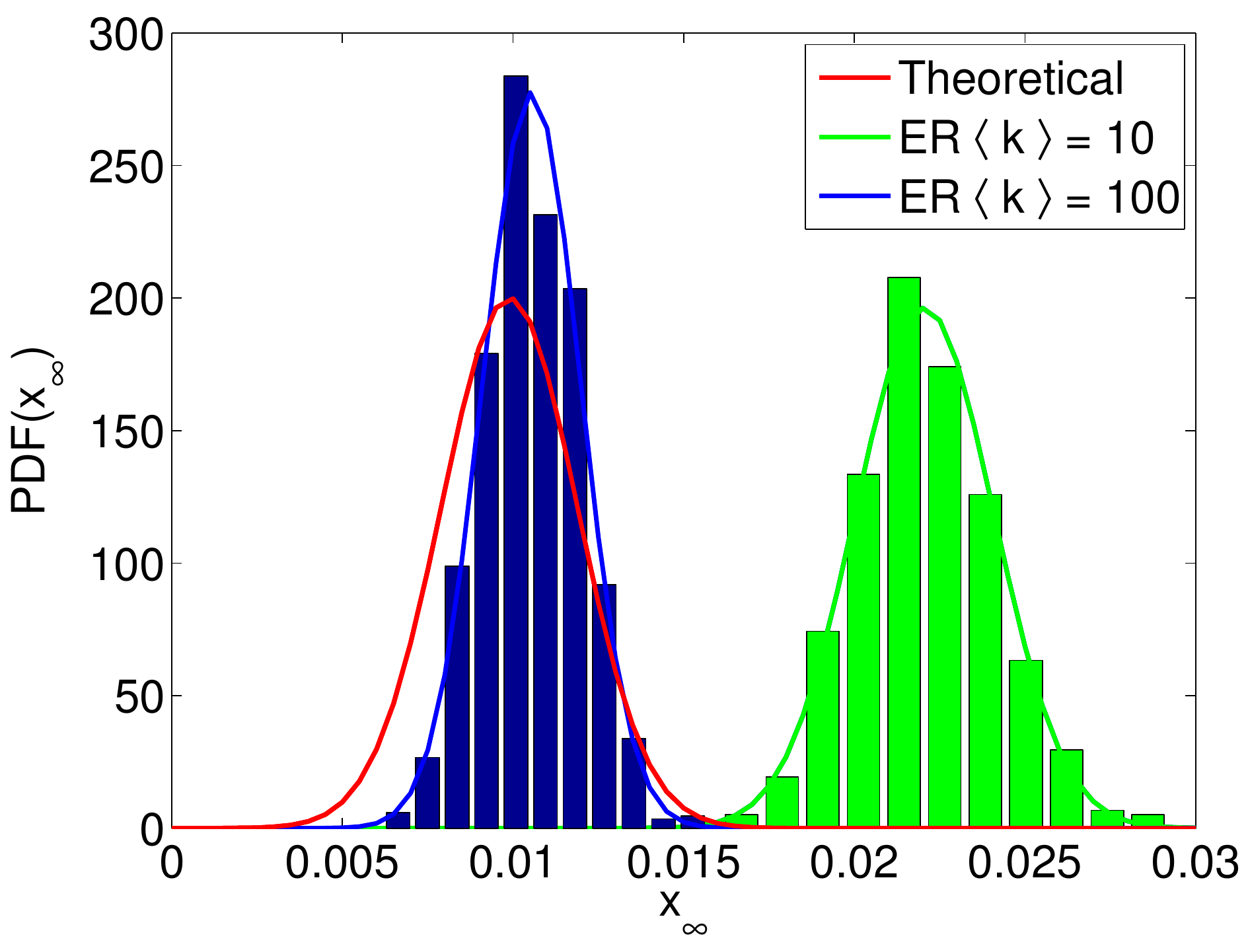}}
\qquad
\subfigure[]{\includegraphics[width=0.45\linewidth]{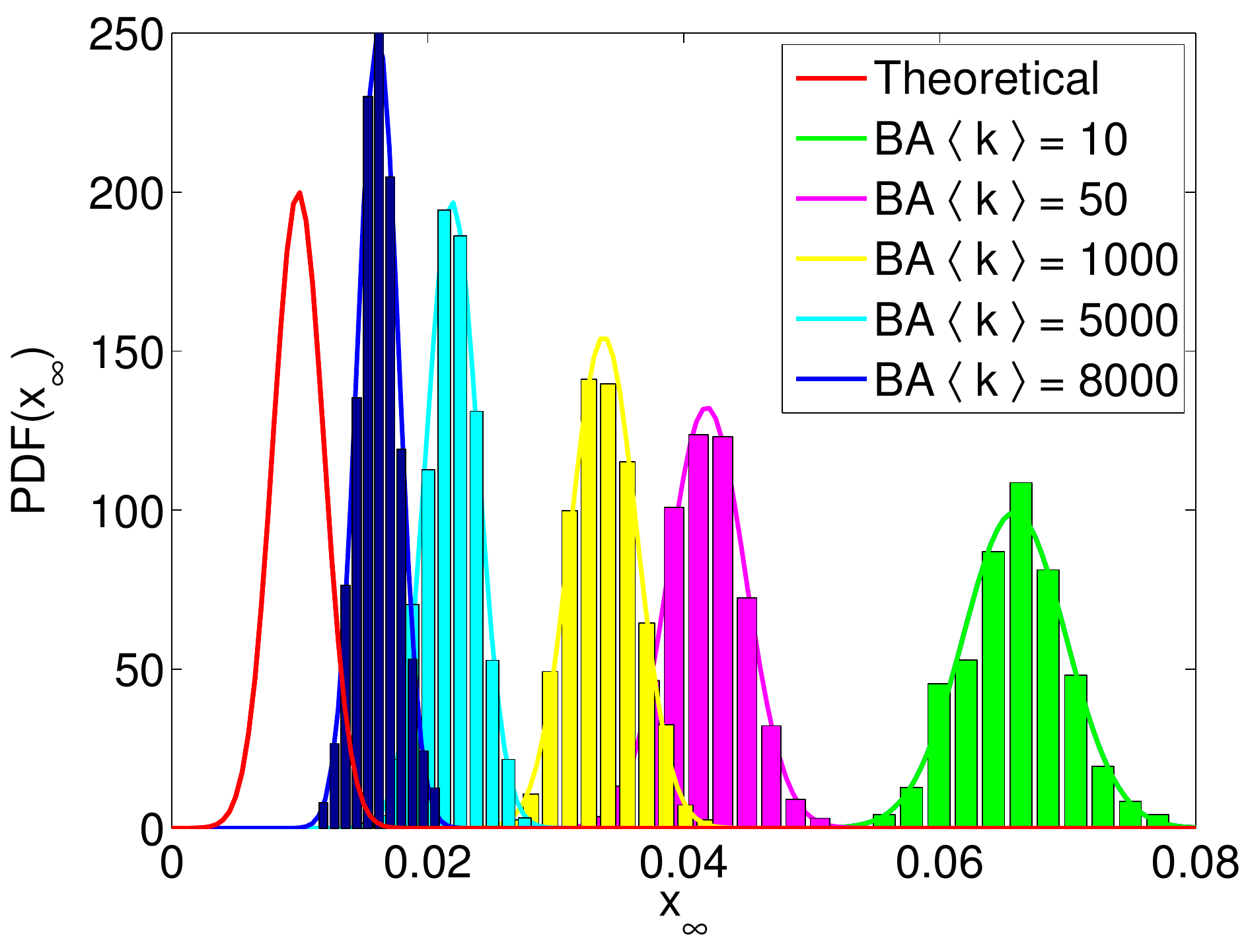}}
\end{center}
\caption{Distribution of the fraction of ignorants considering 1000 Monte Carlo simulations of the rumor scotching model in networks with $n = 10^4$ nodes generated from the (a) ER and (b) BA network models. The simulations consider  $\lambda = 0.5$, $\alpha = 0.5$ and initial conditions $x_0 = 0.98$, $y_0 = 0.01$ and $z_0 = 0.01$. Theoretical curves, obtained by Eqs.~\eqref{eq:edo},~\eqref{eq:fx} and~\eqref{CLT}, are in red.}
\label{fig:tlc}
\end{figure*}

In order to verify the behavior of the rumor scotching model on complex networks, we evaluate networks generated by random graphs of the Erd\H{o}s and R\'{e}nyi (ER) and scale-free networks of Barab\'{a}si and Albert (BA). Figure~\ref{fig:tlc} shows the distribution of the final fraction of ignorants considering 1000 Monte Carlo simulations of the rumor scotching model in networks with $n = 10^4$ vertices generated from the ER and BA models. The theoretical results for the homogeneously mixing populations, obtained from Eqs.~\eqref{eq:edo},~\eqref{eq:fx} and~\eqref{CLT}, are also shown. In ER networks, the distribution converges to the theoretical results as the network becomes denser. In this way, even in sparse networks, $\langle k \rangle = 100$,  the results are close to the mean-field predictions. On the other hand, the convergence of scale-free networks to the theoretical results does not occur even for $\langle k \rangle = 8000$ due to their high level of heterogeneity. 

The system of Eqs.~\ref{eq:edo} that describes the evolution of rumor dynamics on homogeneous populations can characterize the same dynamics in random regular networks if we consider $\lambda = \langle k \rangle \lambda'$ and $\alpha = \langle k \rangle \alpha'$. In this case, the probabilities of spreading and scotching the rumor depend on the number of connections, but the solution of the system of equations does not change. Since random networks present an exponential decay near the mean degree, their dynamical behavior is similar to the mean-field predictions. On the other hand, this approximation is not accurate for scale-free networks, because they do not present a typical degree and the second-moment of their degree distribution diverges for $2 < \gamma \leq 3$ as $n\rightarrow \infty$. Therefore, the homogeneous mixing assumption is suitable only for ER networks.

\begin{figure*}[!ht]
\begin{center}
\subfigure[][$x_0 = 0.98$, $y_0 = 0.005$ and $z_0 = 0.015$.]{\includegraphics[width=0.24\linewidth]{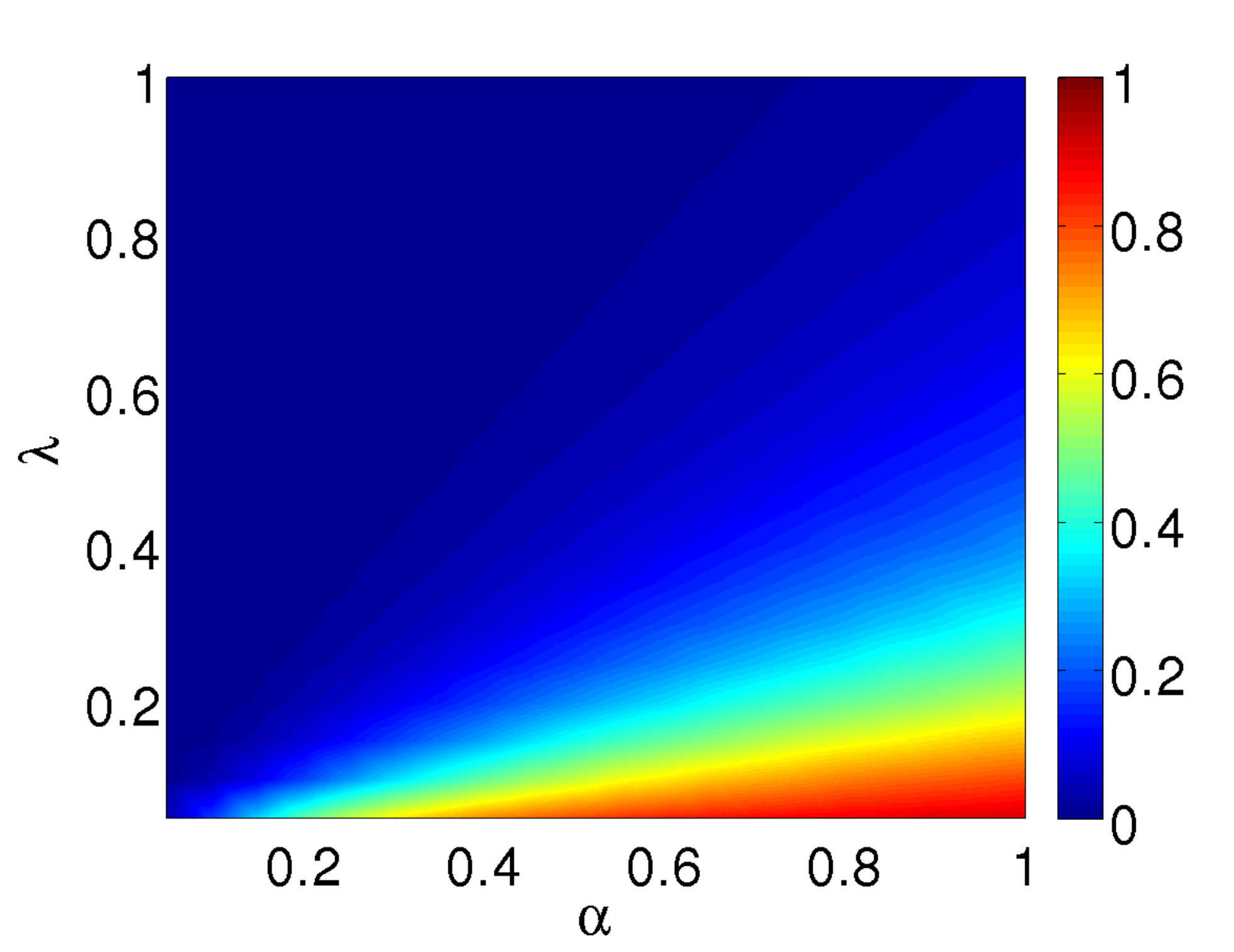}}
\subfigure[][$x_0 = 0.98$, $y_0 = 0.015$ and $z_0 = 0.005$.]{\includegraphics[width=0.24\linewidth]{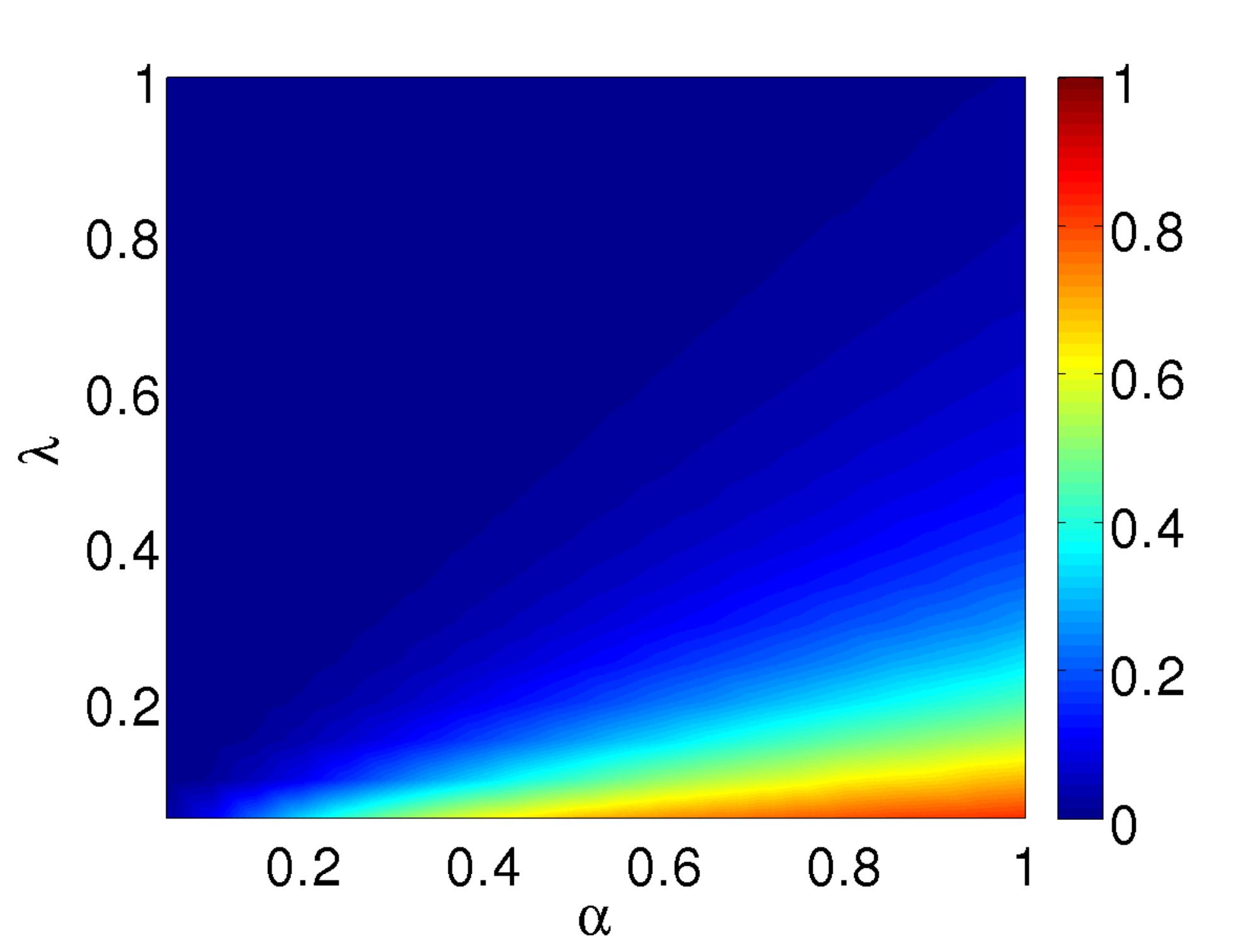}}
\subfigure[][$x_0 = 0.98$, $y_0 = 0.01$ and $z_0 = 0.01$.]{\includegraphics[width=0.24\linewidth]{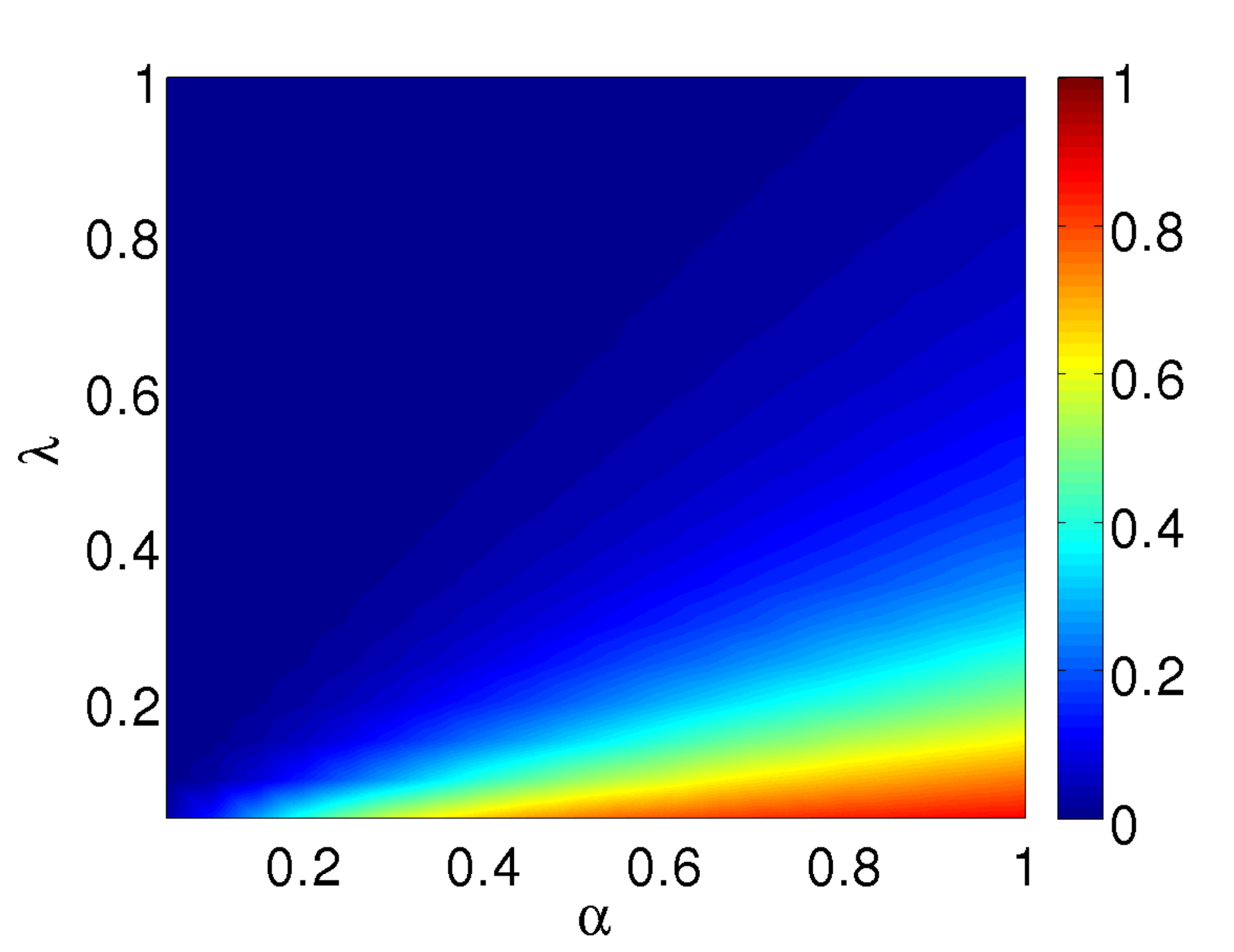}}
\subfigure[][$x_0 = 0.9$, $y_0 = 0.05$ and $z_0 = 0.05$.]{\includegraphics[width=0.24\linewidth]{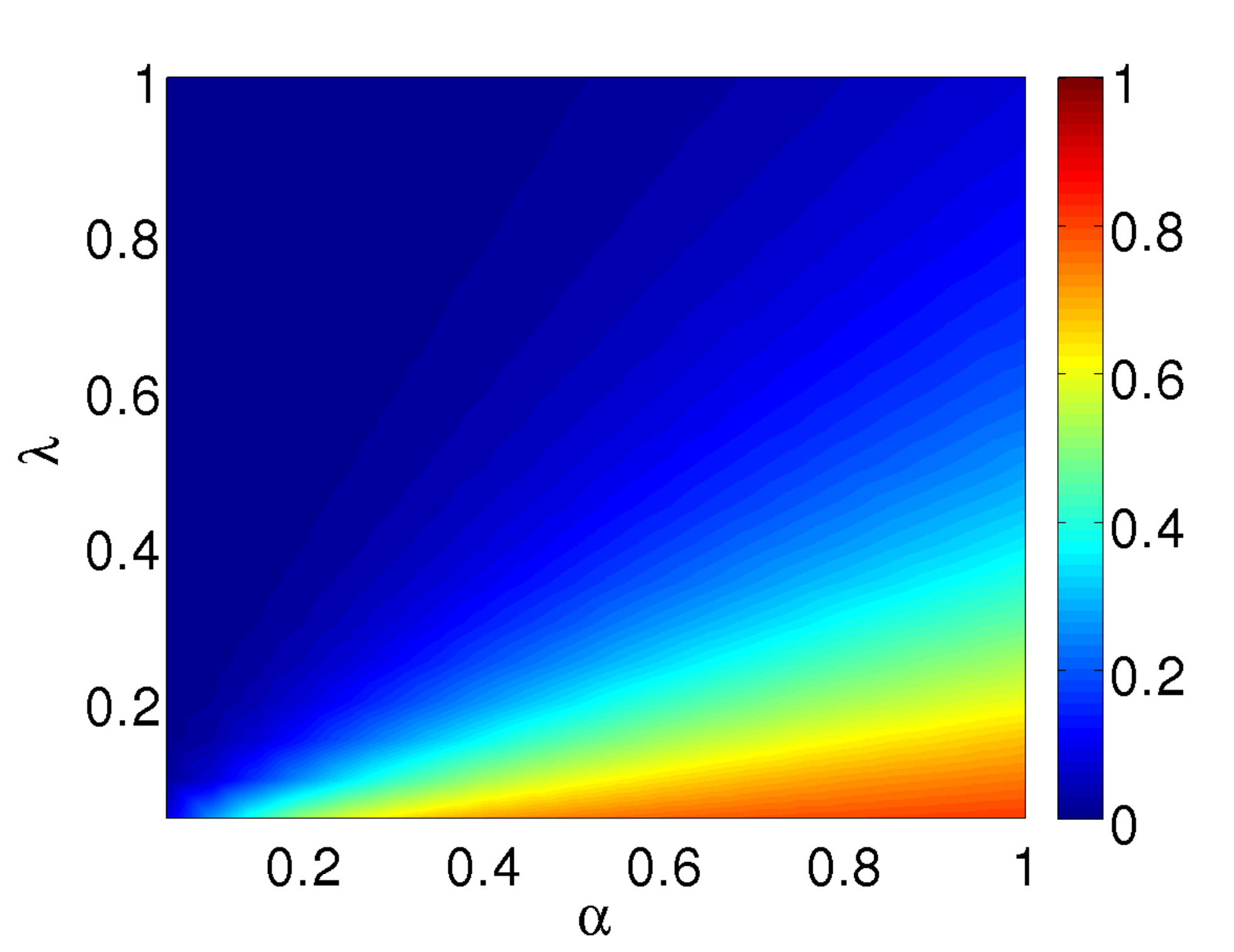}}

\subfigure[][$x_0 = 0.98$, $y_0 = 0.005$ and $z_0 = 0.015$.]{\includegraphics[width=0.24\linewidth]{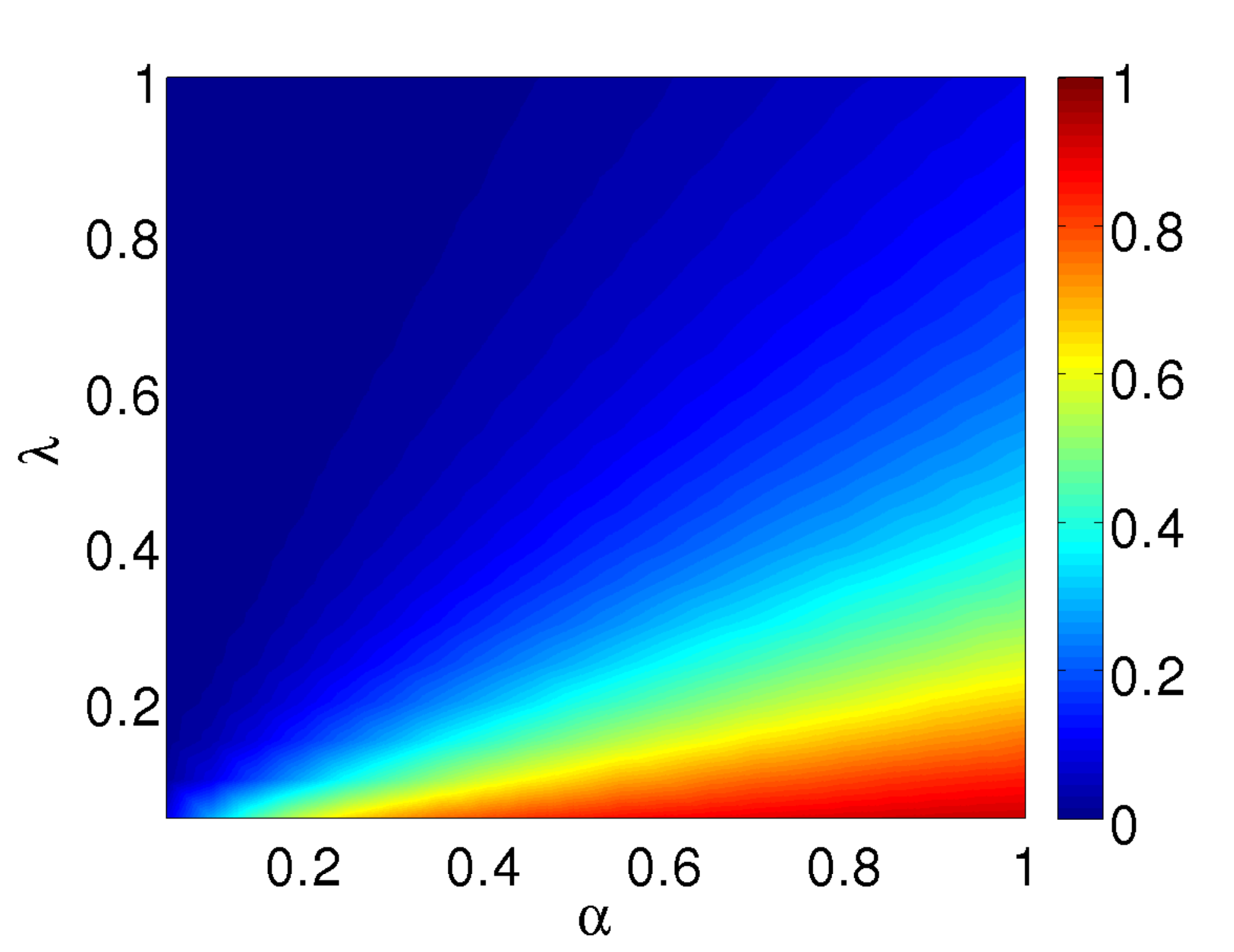}}
\subfigure[][$x_0 = 0.98$, $y_0 = 0.015$ and $z_0 = 0.005$.]{\includegraphics[width=0.24\linewidth]{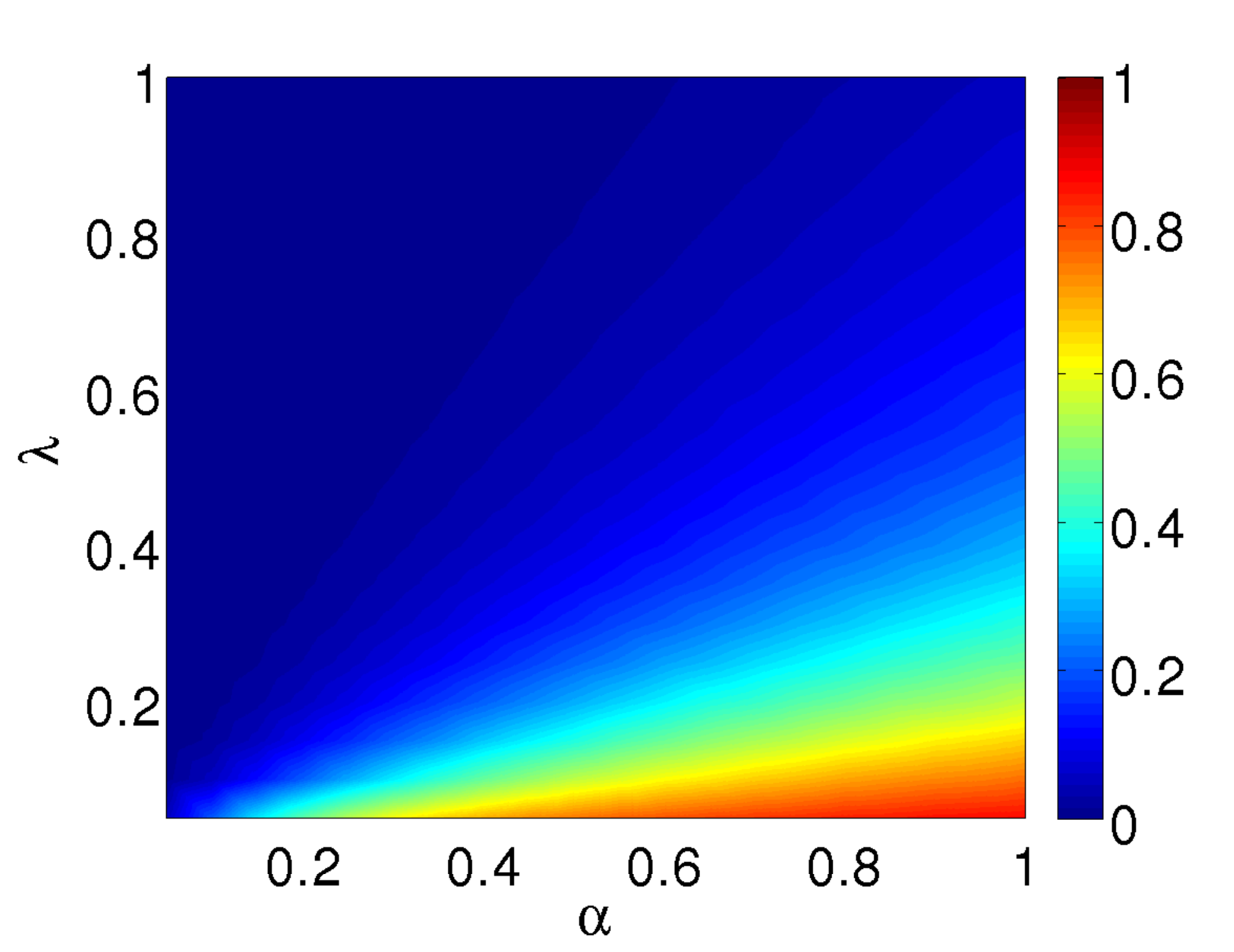}}
\subfigure[][$x_0 = 0.98$, $y_0 = 0.01$ and $z_0 = 0.01$.]{\includegraphics[width=0.24\linewidth]{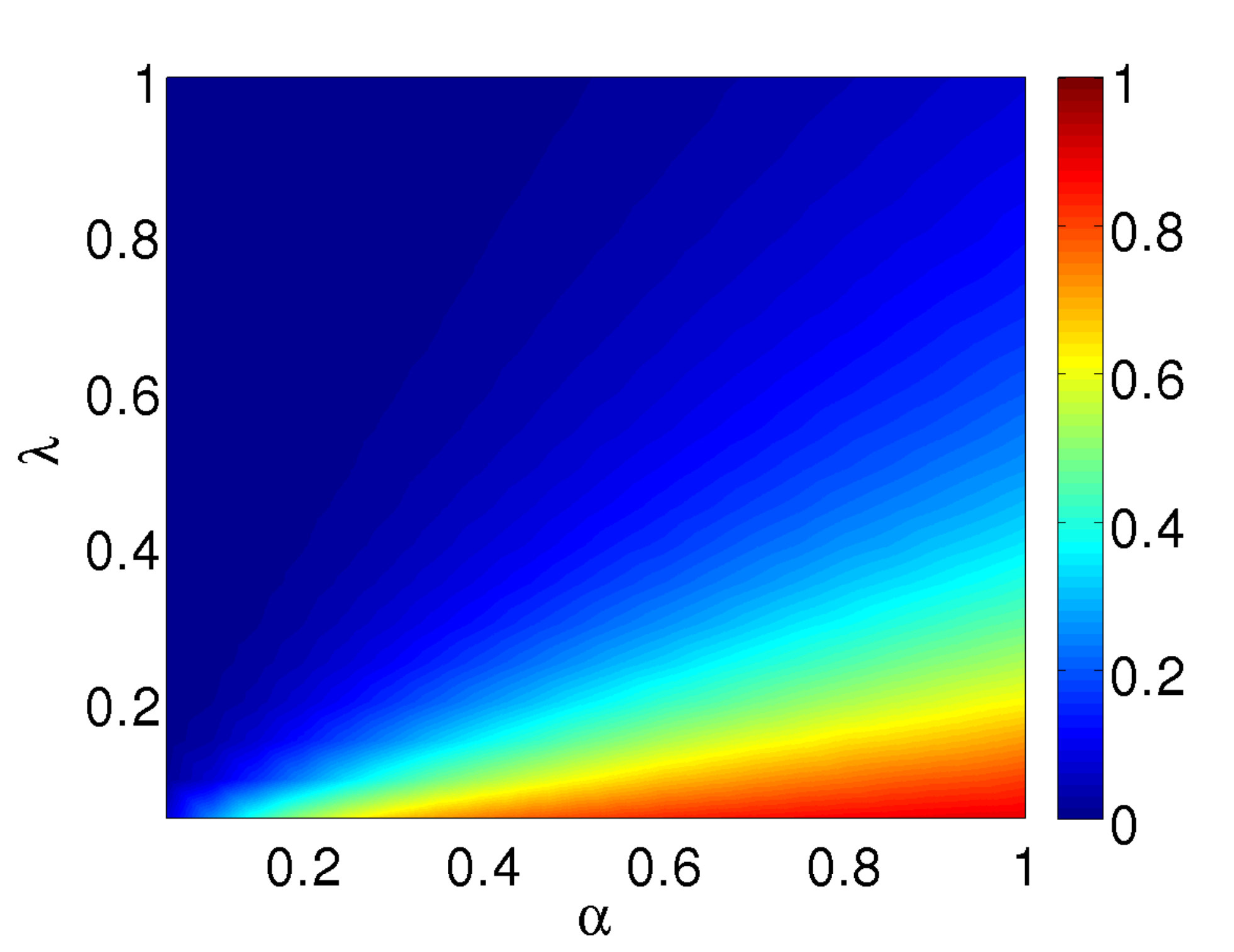}}
\subfigure[][$x_0 = 0.9$, $y_0 = 0.05$ and $z_0 = 0.05$.]{\includegraphics[width=0.24\linewidth]{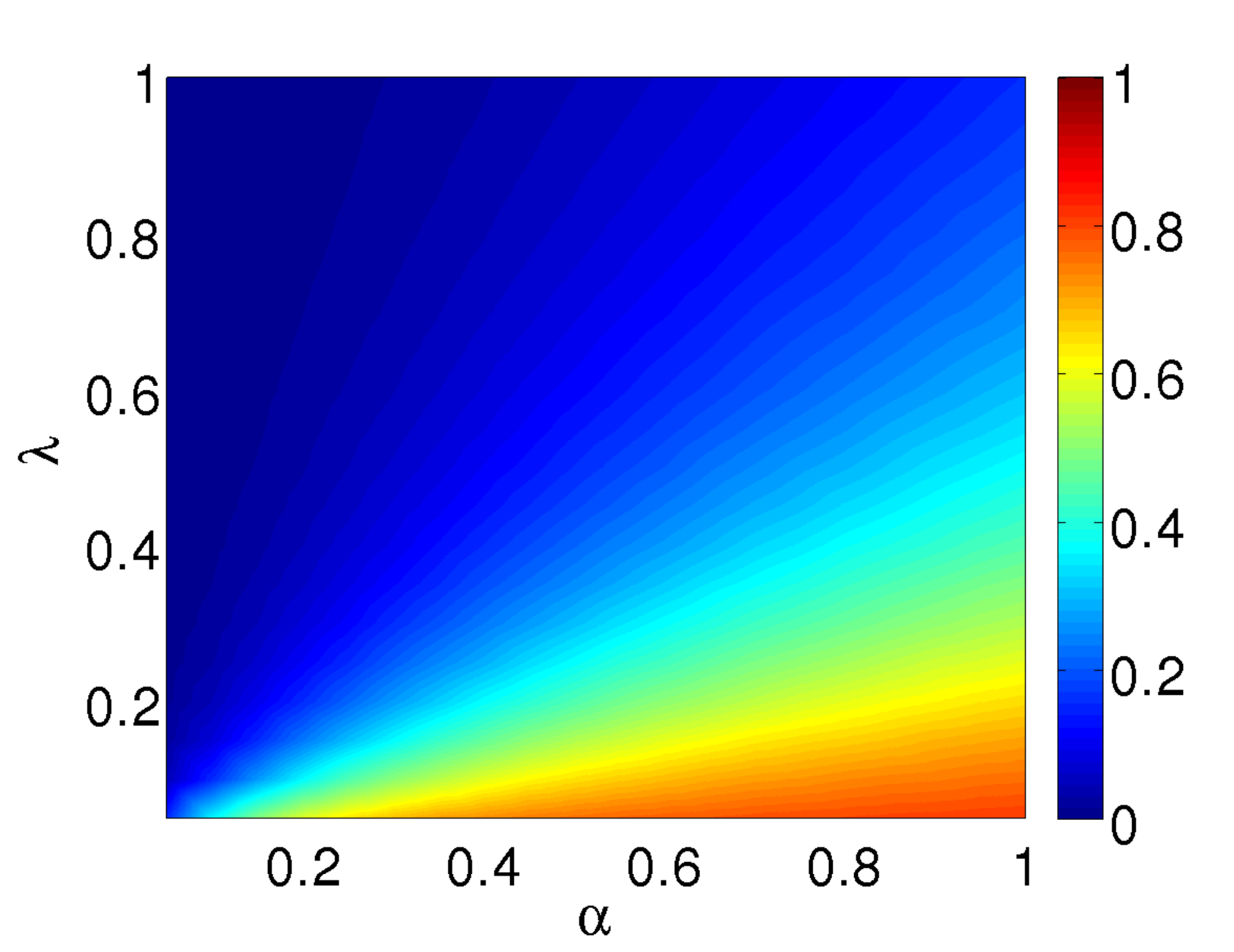}}

\end{center}
\caption{Fraction of ignorants (given by color intensities) according to the rates $\alpha$ and $\lambda$ for different initial conditions considering ER, from (a) to (d), and BA network models, from (e) to (h). Networks with $n = 10^4$ and $\langle k \rangle \approx 8$ are considered. Every point is as an average over 50 simulations.}
\label{fig:initial}
\end{figure*}

Figure~\ref{fig:initial} shows the Monte Carlo simulation results as a function of the parameters $\alpha$ and $\lambda$ for different initial conditions. The simulation considers every pair of parameters, $\lambda$ and $\alpha$, starting from $\lambda = \alpha =$0.05 and incrementing them with steps of 0.05 until reach the unity. In the rumor spreading dynamics, the role played by the stiflers is completely different from the recovered individuals in epidemic spreading. Note that stifler and recovered are absorbing states. However, in the disease spreading, the recovered individuals do not participate on the dynamics and are completely excluded from the interactions, whereas in our model, stiflers are active and try to scotch the rumor to the spreaders.

The number of connections of the initial propagators influences the spread of disease~\cite{Kitsak010, PhysRevE.90.032812}, but does not impact the rumor dynamics~\cite{Borge012}. We investigate if the number of connections of the initial set of spreaders and stiflers affects the evolution of the rumor process with scotching in Barab\'asi-Albert scale-free networks. In a first configuration, the initial state of the hubs is set as spreaders and stiflers are distributed uniformly in the remaining of the network. In another case, stiflers are the main hubs and spreaders are distributed uniformly. In both cases, we verify that the final fraction of ignorants is the same as in completely uniform distribution of spreader and stifler states (see Figure~\ref{fig:initial} (e) -- (h)). Therefore, we infer that the degree of the initial spreaders and stiflers does not influence the final fraction of ignorants.

\begin{figure}[!t]
\begin{center}
\subfigure[][ER network.]{\includegraphics[width=0.98\linewidth]{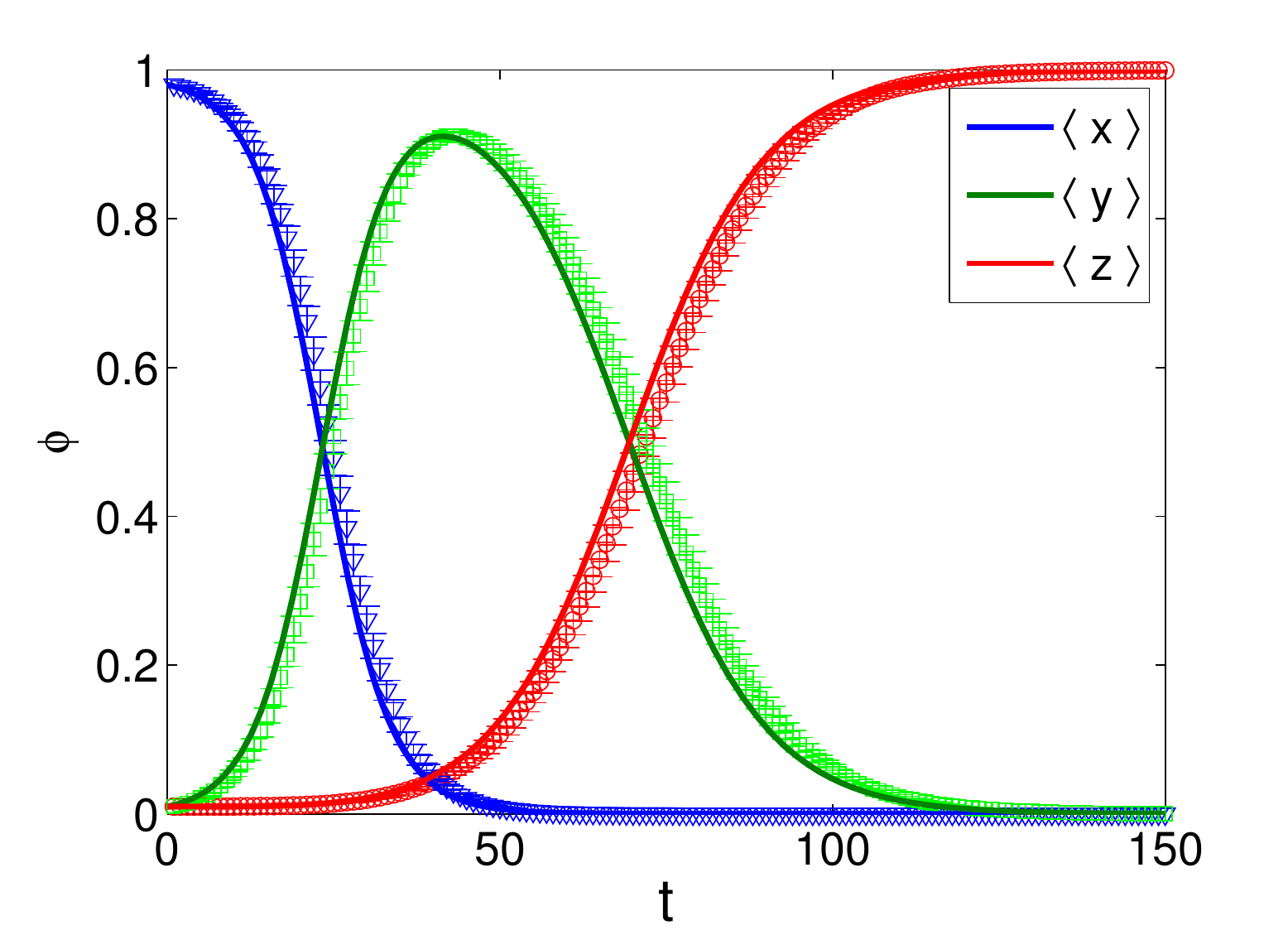}}
\subfigure[][BA network.]{\includegraphics[width=0.98\linewidth]{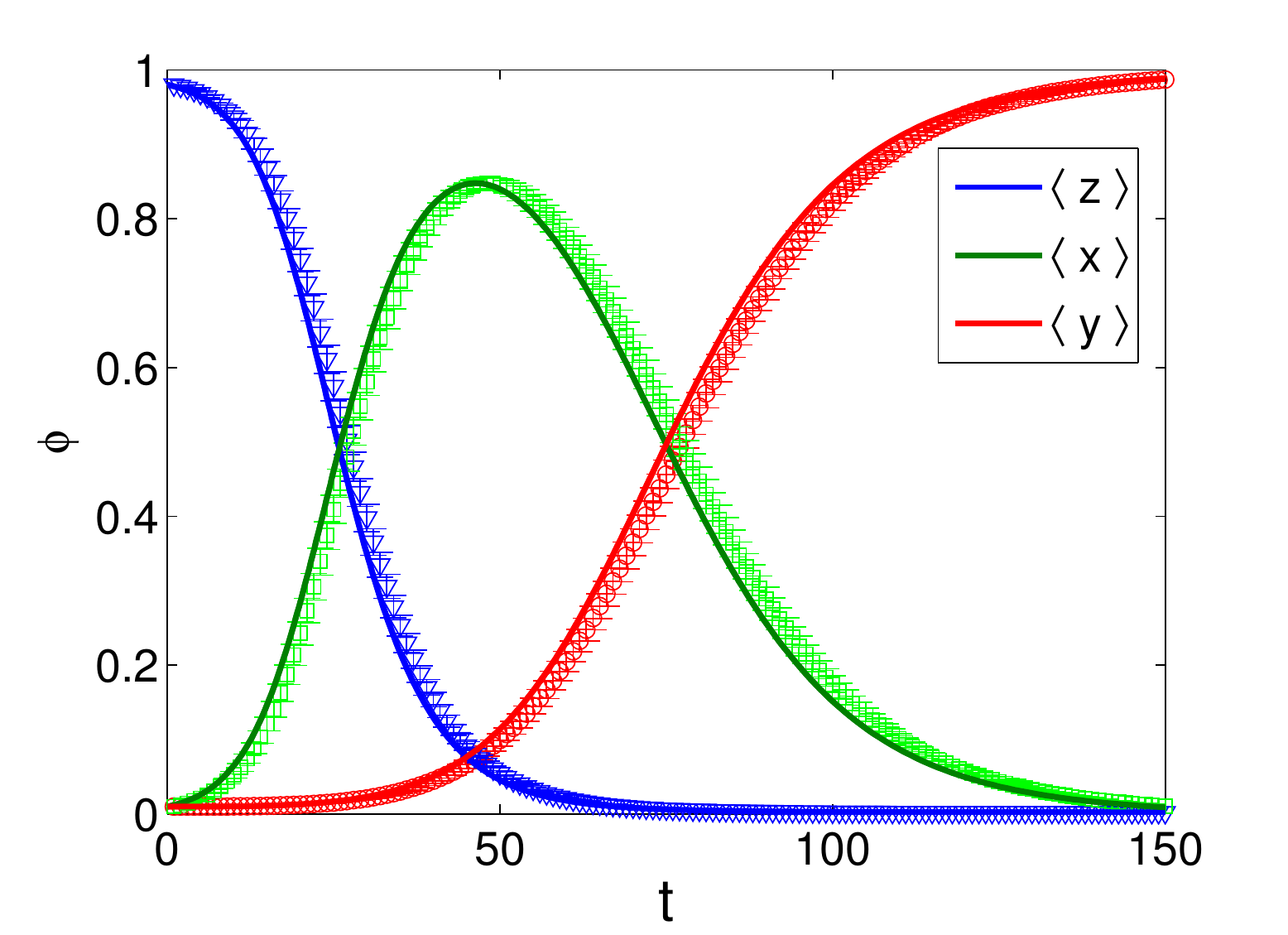}}
\end{center}
\caption{Comparison of the Monte Carlo simulations and the solution of the nodal time evolution differential equations, equations~\ref{eq:network_edo}. The continuous curves are the numerical solution of the differential equations~\ref{eq:network_edo}, while the symbols are the Monte Carlo simulations with its respective standard deviation. Every point is as an average over 50 simulations. In (a) an ER network while in (b) a BA network. Both with $n = 10^4$ nodes and $\langle k \rangle \approx 100$. Moreover, the initial conditions are $x_0 = 0.98$, $y_0 = 0.01$ and $z_0 = 0.01$.}
\label{fig:MC_MK}
\end{figure}

Figure~\ref{fig:MC_MK} shows numerical solutions of equation~\eqref{eq:network_edo} and the Monte Carlo simulations for ER and BA networks. Regarding the simulations, Figures~\ref{fig:MC_MK}(a) and ~\ref{fig:MC_MK}(b) correspond to the average behavior of the variables shown in Figures~\ref{fig:ER} and~\ref{fig:BA}. We can see that the maximum fraction of spreaders occurring in BA networks is lower than in ER networks. This happens because most of the vertices in BA networks are lowly connected (due to the power-law degree distribution). Moreover, we can see that the variance decays over time, which is a consequence of the presence of an absorbing state. In addition we also find that for sparser networks the matching is less accurate (results not shown).

\section{Conclusions}

The modeling of rumor-like mechanisms is fundamental to understand many phenomena in society and on-line communities, such as viral marketing or social unrest. Many works have investigated the dynamics of rumor propagation in complete graphs (e.g.~\cite{daley/kendall/1964}) and complex structures (e.g.~\cite{moreno/nekovee/pacheco/2004}). The models considered so far assume that spreaders try to propagate the information, whereas stiflers are not active. Here, we propose a new model in which stiflers try to scotch the rumor to the spreader agents. We develop an analytical treatment to determine how the fraction of ignorants behaves asymptotically in finite populations by taking into account the homogeneous mixing assumption. We perform Monte Carlo simulations of the stochastic model on Erd\H os-R\'enyi random graphs and Barab\'asi-Albert scale-free networks. The results obtained for homogeneously mixing populations can be used to approximate the case of random networks, but are not suitable for scale-free networks, due to their highly heterogeneous organization. The influence of the number of connections of the initial spreaders and stiflers is also addressed. We verify that the choice of hubs as spreaders or stiflers has no influence on the final fraction of ignorants. 

The study performed here can be extended by considering additional network models, such as small-world or spatial networks. The influence of network properties, such as assortativity and community organization can also be analyzed in our model. In addition, strategies to maximize the range of the rumor when the scotching is present can also be developed. The influence of the fraction of stiflers on the final fraction of ignorant vertices is another property that deserves to be investigated. 

\section{Acknowledgements}

PMR acknowledges FAPESP (grant 2013/03898-8) and CNPq (grant 479313/2012-1) for financial support. FAR acknowledges CNPq (grant 305940/2010-4), FAPESP (grants 2011/50761-2 and 2013/26416-9) and NAP eScience - PRP - USP for financial support. GFA acknowledges FAPESP for the sponsorship provided. EL acknowledges CNPq (grant 303872/2012-8), FAPESP (grant 2012/22673-4) and FAEPEX - UNICAMP for financial support.

\section{Appendix}

In this section we describe the main steps behind the proofs of our results presented for homogeneously mixing populations. Similar arguments have been applied for stochastic rumor and epidemic models 
\cite{Ball/Britton/2007,lebensztayn/machado/rodriguez/2011a,lebensztayn/machado/rodriguez/2011b} and we include them for the sake of completeness. First we note that according to Theorem 11.2.1 of \cite{Ethier2009} we have that, on a suitable
probability space, $\tilde{V}^n(t)/n$ converges to $(x(t),y(t))$ given by \eqref{eq:xinf}, almost surely uniformly on bounded time intervals. Then our results can be obtained as a direct consequence of
Theorem 11.4.1 of \cite{Ethier2009}. To show this we use the notation used there, except for the Gaussian process that we would rather denote by $\mathcal{V} =
(\mathcal{V}_x, \mathcal{V}_y)$. Here  $\varphi(x, y) = y$, and
$$\tau_\infty = \inf \{t: y(t) \leq 0 \} = -\frac{1}{\lambda}\log \left(\frac{x_\infty}{x_0}\right).$$
Moreover, 
\begin{equation}
\label{F:Der_Neg}
\nabla \varphi(v(\tau_\infty)) \cdot F(v(\tau_\infty)) =  y^{\prime} (\tau_\infty) = (\lambda +\alpha) x_\infty - \alpha < 0.
\end{equation}

\subsection{Law of Large Numbers}
\label{sub:LLN}

In order to prove the limit of Eq.~\eqref{conv:probability}, note that $y_0 >0$ and \eqref{F:Der_Neg} imply that  $y(\tau_\infty - \epsilon) > 0$ and $y(\tau_\infty + \epsilon) < 0$ for $0 < \epsilon <
\tau_\infty$. Therefore the almost surely convergence of $\tilde{y}^n(t)$ to $y(t)$ uniformly on bounded intervals implies that 
\begin{equation}
\label{F:Conv_tf}
\lim_{n \to \infty} \, \tilde \tau^{(n)} = \tau_\infty \quad \text{a.s.}
\end{equation}
When $ y_0 = 0 $ and $ x_0 > \rho z_0 $, this result is also valid because $ y^{\prime}(0) > 0$ and \eqref{F:Der_Neg} still holds.
On the other hand, if $ y_0 = 0 $ and $ x_0 \leq \rho z_0$, then $y(t) < 0$ for all $t > 0$, and again the almost sure convergence of~$\tilde{Y}^n(t)/n$ to~$y$ uniformly on
bounded intervals yields that $ \lim_{n \to \infty} \tilde \tau^{(n)} = 0 = \tau_\infty $ almost surely.
Therefore, as $\tilde{X}^n(t)/n$ converges to $x(t)$ almost surely, we obtain the LLN from \eqref{F:Conv_tf} and the fact that $X^{(n)}(\tau^{(n)}) = \tilde{X}^{(n)}(\tilde{\tau}^{(n)})$.

\subsection{Central Limit Theorem}
\label{sub:CLT}

Now, we show the arguments to prove the central limit in Eq.~\eqref{CLT}. From Theorem 11.4.1 of \cite{Ethier2009} we have that if, $y_0 >0$ or $y_0 =0$ and $x_0 > \rho z_0$, then
$$\sqrt{n} \, ( n^{-1} \tilde{X}^n(\tilde \tau^{(n)}) - x_\infty)$$
converges in distribution as $n \to \infty$ to
\begin{equation}
\label{F:LD}
\mathcal{V}_x(\tau_\infty) + \frac{x_\infty}{(1+\delta)x_\infty -\delta} \, \mathcal{V}_y(\tau_\infty).
\end{equation}
The resulting normal distribution has mean zero, so, to complete the proof of CLT, we need to calculate the corresponding variance. To compute the covariance matrix 
$\text{Cov}(\mathcal{V}(\tau_\infty), \mathcal{V}(\tau_\infty))$, we use Eq. (2.21) from \cite[Chap. 10]{Ethier2009} which translates to
\begin{equation}
\label{F: Covariance}
\text{Cov}(\mathcal{V}(t), \mathcal{V}(t)) = \int_0^{t} \Phi(t, s) \, G(x(s), y(s)) \, {[\Phi(t, s)]}^T \, ds.
\end{equation}
In our case,
$$G(x, y) = 
\begin{pmatrix}
\lambda x & -\lambda x \\
-\lambda x & (\lambda - \alpha)x - \alpha y+ \alpha
\end{pmatrix}
$$
and
$$ 
\Phi(t, s) = 
\begin{pmatrix} 
e^{-\lambda(t-s)} & 0 \\ 
e^{\alpha(t-s)} - e^{-\lambda(t-s)} & e^{\alpha(t-s)}
\end{pmatrix},
$$
thus we obtain that $\text{Cov}(\mathcal{V}(\tau_\infty), \mathcal{V}(\tau_\infty))$ is given by
$$
\left(\begin{array}{cc}
x_{\infty}-\dfrac{x_{\infty}^2}{x_0} & \dfrac{x_{\infty} (x_{\infty}-x_0)}{x_0} \\[0.4cm]
\dfrac{x_{\infty} (x_{\infty}-x_0)}{x_0} & 2 x_{\infty} -1 + \dfrac{(x_{\infty}-1)^2}{z_0}-\dfrac{x_{\infty}^2}{x_0}
\end{array}
\right).
$$
We get the closed formula \eqref{variancia} for the asymptotic variance by using last expression and properties of variance.

\bibliographystyle{apsrev}
\bibliography{bibliography}

\end{document}